\documentclass[11pt]{article}
\pdfoutput=1
\usepackage{arxiv}
\usepackage{microtype}%if unwanted, comment out or use option "draft"
\usepackage[table]{xcolor}
\usepackage[T1]{fontenc}
\usepackage{amsthm}
\usepackage{amssymb}
\usepackage[english]{babel}
\usepackage[utf8]{inputenc}
\usepackage{amsmath}
\usepackage{amsfonts}
\usepackage{graphicx}
\usepackage[colorinlistoftodos]{todonotes}
\usepackage{mathtools}
\usepackage{enumitem}
\usepackage{thmtools}
\usepackage{fullpage}
\usepackage{booktabs}
\usepackage{bbm}
\usepackage[normalem]{ulem} % m.in. przekreślenie tekstu
\usepackage[hidelinks]{hyperref}
\usepackage{multirow}
\newcommand{\N}{\mathbb{N}}
\newcommand{\R}{\mathbb{R}}

\newcommand{\A}{\mathbf{A}}

\newcommand{\vol}{\textrm{vol}}
\newcommand{\ABCD}{\textbf{ABCD}}

\theoremstyle{plain}

\usepackage{algorithm}
\usepackage{algpseudocode}
\usepackage{subfig}

\title{Modularity Based Community Detection in Hypergraphs} 

\author{
Bogumi\l{} Kami\'nski\thanks{Decision Analysis and Support Unit, SGH Warsaw School of Economics, Warsaw, Poland; e-mail: \texttt{bogumil.kaminski@sgh.waw.pl}}
\And
Pawe\l{} Misiorek\thanks{Institute of Computer Sciences, Poznan University of Technology, Poznan, Poland; e-mail: \texttt{pawel.misiorek@put.poznan.pl}}
\And 
Pawe\l{}~Pra\l{}at\thanks{Department of Mathematics, Toronto Metropolitan University, Toronto, ON, Canada; e-mail: \texttt{pralat@torontomu.ca}}
\And
Fran\c{c}ois Th\'eberge\thanks{Tutte Institute for Mathematics and Computing, Ottawa, ON, Canada; email: \texttt{theberge@ieee.org}}
}

\begin{document}

\maketitle

\begin{abstract}
In this paper, we propose a scalable community detection algorithm using hypergraph modularity function, \textbf{h--Louvain}. It is an adaptation of the classical \textbf{Louvain} algorithm in the context of hypergraphs. We observe that a direct application of the \textbf{Louvain} algorithm to optimize the hypergraph modularity function often fails to find meaningful communities. We propose a solution to this issue by adjusting the initial stage of the algorithm via carefully and dynamically tuned linear combination of the graph modularity function of the corresponding two-section graph and the desired hypergraph modularity function. The process is guided by Bayesian optimization of the hyper-parameters of the proposed procedure. Various experiments on synthetic as well as real-world networks are performed showing that this process yields improved results in various regimes.
\end{abstract}

%%%%%%%%%%%%%%%%%%%%%%%%%%%%%%%%%%%%%%%%%%%%%%%%%%%%%%%%%%%
\section{Introduction}
%%%%%%%%%%%%%%%%%%%%%%%%%%%%%%%%%%%%%%%%%%%%%%%%%%%%%%%%%%%

Many networks that are currently modelled as graphs would be more accurately modelled as hypergraphs. This includes the collaboration network~\cite{benson2018simplicial} in which nodes correspond to researchers and hyperedges correspond to papers that consist of nodes associated with researchers that co-author a given paper. Social events may include more than two people which is not equivalent to social interactions among all pairs of people participating in the event. Hypergraphs have shown promise in modeling systems such as protein complexes and metabolic reactions~\cite{feng2021hypergraph}. Another natural examples are co-purchases hypergraphs but there are plenty of other real-world hypergraphs. 

After many years of intense research using graph theory in modelling and mining complex networks~\cite{easley2010networks,jackson2010social,kaminski2021mining,newman2018networks}, hypergraphs start gaining considerable traction~\cite{battiston2020networks,benson2018simplicial,benson2021higher,benson2016higher}. Many higher-order network data is being collected in recent years (see, for example,~\cite{benson2018simplicial}). It has became clear to both researchers and practitioners that dyadic relationships are insufficient in many real-world scenarios. Higher-order network analysis, using the ideas of hypergraphs, simplicial complexes, multilinear and tensor algebra, and more, is needed to study complex systems and to make an impact across many important applications~\cite{benson2021higher,lambiotte2018understanding,tian2024higher,lee2024survey}. Indeed, the inherent expressiveness of hypergraphs has led to their applications across a diverse range of fields such as recommendation systems~\cite{xia2021self}, computer vision~\cite{liao2021hypergraph}, natural language processing~\cite{ding2020more}, social network analysis~\cite{matwin2023generative}, financial analysis~\cite{yi2022structure}, bioinformatics~\cite{feng2021hypergraph}, and circuit design~\cite{grzesiak2017hypergraphs}. Standard but important questions in network science are currently being revisited in the context of hypergraphs. However, hypergraphs also create brand new questions which did not have their counterparts for graphs. For example, how hyperedges overlap in empirical hypergraphs~\cite{lee2021hyperedges}? Or how the existing patters in a hypergraph affect the formation of new hyperedges~\cite{juul2022hypergraph}?

\medskip

In this paper we concentrate on the classical problem of \emph{community detection} in networks that can be represented using hypergraphs~\cite{ahn2018hypergraph,benson2015tensor,chien2018community,chodrow2021generative,kaminski2019clustering,kaminski2020community,Kumar2,Kumar1,yin2018higher,yin2017local}. Community detection is a challenging, NP-hard problem even for graphs~\cite{brandes2007modularity,fortunato2010community,newman2006modularity} so obtaining an optimal solution becomes computationally infeasible, even for small networks represented as graphs. Dealing with hypergraphs is clearly much more difficult so, despite the fact that currently there is a vivid discussion around hypergraphs, the theory and tools are still not sufficiently developed to tackle this problem directly within this context. Indeed, researchers and practitioners, due to lack of proper solutions for hypergraphs, often create the 2-section graph of a hypergraph of interest (that is, replace each hyperedge with a clique, a process known also as clique expansion). Given the 2-section graph representation, we can directly apply some graph clustering algorithm such as \textbf{Louvain}~\cite{blondel2008fast} and \textbf{Leiden}~\cite{traag2019louvain}. Another approach is to perform agglomerative clustering via some definition of distance between nodes, such as the derivative graph defined in Contreras-Aso et al.~\cite{contreras-aso2023}, and then select the partition that maximizes the 2-section graph modularity. However, with the 2-section graph, one clearly loses some information about hyperedges of size greater than two. In the experiments presented in Section~\ref{sec:experiments}, we use the \textbf{Louvain} algorithm on the 2-section graph representations as our basis of comparison for hypergraph-based algorithms. 

As mentioned earlier, there are some recent attempts to deal with hypergraphs in the context of clustering. For example, Kumar et al.~\cite{Kumar2,Kumar1} still reduce the problem to graphs but use original hypergraphs to iteratively adjust weights to encourage some hyperedges to be included in some cluster but discourage other ones (this process can be viewed as separating signal from noise). In Chodrow et al.~\cite{chodrow2021generative}, a hypergraph stochastic block model is defined, leading to a Louvain-type clustering algorithm, in particular, for the ``all or nothing'' regime (\textbf{AON}), where edges must have all nodes from the same community to improve the objective function. We provide more details about these two algorithms at the beginning of Section~\ref{sec:experiments}. 

Many of the successful graph clustering algorithms use the modularity function to benchmark partitions to guide the associated optimization heuristics. Two widely used algorithms from this family are the {\bf Louvain} and {\bf Leiden} algorithms mentioned earlier. Based on its spectacular success, a number of extensions of the classical graph modularity function to hypergraphs are proposed~\cite{kaminski2019clustering,kaminski2020community} that can potentially be used by true hypergraph algorithms. In this paper, we concentrate on this approach. 

\medskip

Unfortunately, there are many ways such extension of the modularity function to hypergraphs can be done, depending on how often nodes in one community share hyperedges with nodes from other communities. We believe that the underlying process that governs \emph{pureness} of community hyperedge is something that varies between networks at hand and also potentially depends on the hyperedge sizes. Let us come back to the collaboration network we discussed earlier. Hyperedges associated with papers written by mathematicians might be more homogeneous and smaller in comparison with those written by medical doctors who tend to work in large and multidisciplinary teams. Moreover, in general, papers with a large number of co-authors tend to be less homogeneous, and other patterns can be identified~\cite{juul2022hypergraph}. The algorithm we propose in this paper, \textbf{h--Louvain}, is flexible and can use any of such hypergraph modularity function. In other words, there is no unique way of extending the concept of modularity from graphs to hypergraphs. For this reason we consider a family of such extensions parametrized by the user's preference over homogeneity of within-community hyperedges. At the same time we recognize that there can be situations in which it might not be clear for a user what homogeneity level is desired. Therefore, in Section~\ref{sec:modularity_selection} we provide some suggestions to help the user to make the right choice.

A significant challenge in optimizing modularity functions is that these objective functions have their domains defined over all partitions of the set of nodes and they are known to be extremely difficult to optimize. As already mentioned, one of the most popular and efficient heuristic methods for modularity optimization for graphs is the \textbf{Louvain} algorithm~\cite{blondel2008fast}. In this paper, we show how this algorithm can be adapted to optimize hypergraph modularity. One of the main challenges is the fact that, when hyperedges of size two (edges) or three are not present in the hypergraph, then the \textbf{Louvain} algorithm immediately gets stuck in its local minimum. 
Moreover, even if there are a few hyperedges of size two or three, the algorithm may still get stuck almost immediately, and yield a solution that is heavily biased toward small edges. Hence, in such situations, one cannot simply start optimizing the hypergraph modularity right from the beginning. More importantly, we observe that even if hyperedges of size two are present in the hypergraph, the algorithm often converges to a local optimum that is of low quality. In order to address these two problems, we propose a method that works reasonably well in practice in which we optimize a weighted average of the 2-section graph modularity function and the hypergraph modularity function. For that we adjust the \textbf{Louvain} algorithm in such a way that the weight of the hypergraph modularity function increases during the optimization process. The pace of this weight change is governed by two hyperparameters of the procedure, which we tune using Bayesian optimization.

\medskip

The paper is structured as follows. We first introduce the necessary notation; in particular, we state the definitions of graph and hypergraph modularity functions (Section~\ref{sec:modularity}). Synthetic as well as real-world hypergraphs that are used in our experiments are introduced in Section~\ref{sec:hypergraphs}. Section~\ref{sec:algorithm} is devoted to explain details behind the proposed algorithm, \textbf{h--Louvain}. First, we discuss the classical \textbf{Louvain} algorithm for graphs (Subsection~\ref{ssec:Louvain}) and explain why it is difficult to adjust it to directly optimize hypergraph modularity (Subsection~\ref{ssec:challenges}). Following this, we describe our solution that is considering a linear combination of the 2-section graph modularity and the hypergraph modularity as objective function (Subsection~\ref{subsec:algorithm}), and explain its implementation challenges (Subection~\ref{ssec:parameters}). In particular, the main challenge is to tune the two hyperparameters responsible for the speed of convergence to the hypergraph modularity function. To find a ``sweet spot'' in an unsupervised way, Bayesian optimization is used (Subsection~\ref{sec:bayesopt}). Section~\ref{sec:experiments} highlights the results of numerical experiments of using the proposed algorithm on synthetic hypergraphs (Subsections~\ref{sec:experiments_h-ABCD} and~\ref{sec:more_challenging_ABCD}) as well as real-world hypergraphs (Subsection~\ref{sec:exp-real_hypergraphs}). We also highlight important implications of the choice of the modularity function to optimize (Subsection~\ref{sec:modularity_selection}). The paper is concluded with a summary of outlooks for further research in this area (Section~\ref{sec:conclusions}). 

Finally, let us mention that this paper is an extended, journal version of the short, proceeding paper~\cite{kaminski2023modularity} that contained some preliminary experiments with a much simpler algorithm. The algorithm as well as notebooks containing all experiments included in this paper can be found on GitHub repository\footnote{\url{https://github.com/pawelwm/h-louvain}}.

%%%%%%%%%%%%%%%%%%%%%%%%%%%%%%%%%%%%%%%%%%%%%%%%%%%%%%%%%%%
\section{Modularity Functions}\label{sec:modularity}
%%%%%%%%%%%%%%%%%%%%%%%%%%%%%%%%%%%%%%%%%%%%%%%%%%%%%%%%%%%

Let us start with some basic definitions. In the hypergraph $H=(V,E)$, each hyperedge $e \in E$ is a multiset of $V$ of any cardinality $d \in \N$ called its size. Multisets in the context of hypergraphs are natural generalization of loops in the context of graphs. Hypergraphs are natural generalization of graphs in We shouldhich edge is a multiset of size two. Even though $H$ does not always contain multisets, it is convenient to allow them as they may appear in the random hypergraph that will be used as the null model to ``benchmark'' the edge contribution component of the modularity function. It will be convenient to partition the hyperedge set $E$ into $\{E_1, E_2, \ldots \}$, where $E_d$ consists of hyperedges of size $d$. As a result, hypergraph $H$ can be expressed as the disjoint union of \emph{$d$-uniform hypergraphs} $H = \bigcup H_d$, where $H_d = (V, E_d)$. As for graphs, $\deg_{H}(v)$ is the degree of node $v$, that is, the number of hyperedges $v$ is a part of (taking into account the fact that hyperedges are multisets). Finally, the volume of a subset of nodes $A \subseteq V$ is $\vol_H(A) = \sum_{v \in A} \deg_H(v)$.

%%%%%%%%%%%%%%%%%%%%%%%%%%%%%%%%%%%%%%%%%%%%%%%%%%%%%%%%%%%
\subsubsection*{Graph Modularity}

The definition of modularity for graphs was first introduced by Newman and Girvan in~\cite{newman2004finding}.  Despite some known issues with this function such as the ``resolution limit'' reported in~\cite{fortunato2007resolution}, many popular algorithms for partitioning nodes of large graphs use it~\cite{clauset2004finding,lancichinetti2011limits,newman2004fast} and perform very well. The two prominent ones from this family are \textbf{Louvain}~\cite{blondel2008fast} and \textbf{Leiden}~\cite{traag2019louvain}. The modularity function favours partitions of the set of nodes of a graph $G$ in which a large proportion of the edges fall entirely within the parts (often called clusters), but benchmarks it against the expected number of edges one would see in those parts in the corresponding Chung-Lu random graph model~\cite{chung2006complex} which generates random graphs with the expected degree sequence following exactly the degree sequence in $G$. 

Formally, for a graph $G=(V,E)$ and a given partition $\A = \{A_1, A_2, \ldots, A_k\}$ of $V$, the \emph{modularity function} is defined as follows:
\begin{eqnarray}
q_G(\A) &=& \sum_{A_i \in \A} \frac{e_G(A_i)}{|E|}  - \sum_{A_i \in \A} \left( \frac{\vol_G(A_i)}{\vol_G(V)} \right)^2, \label{eq:q_G_A}
\end{eqnarray}
where $e_G(A_i)$ is the number of edges in the subgraph of $G$ \emph{induced by} set $A_i$. The first term in~(\ref{eq:q_G_A}), $\sum_{A_i \in \A} e_G(A_i)/|E|$, is called the \emph{edge contribution} and it computes the fraction of edges that fall within one of the parts. The second one, $\sum_{A_i \in \A} (\vol_G(A_i)/\vol_G(V))^2$, is called the \emph{degree tax} and it computes the expected fraction of edges that do the same in the corresponding random graph (the null model). The modularity measures the deviation between the two.

%It is easy to see that $q_G(\A) \le 1$. Also, if $\A = \{V\}$, then $q_G(\A) = 0$, and if $\A = \{ \{v_1\}, \{v_2\}, \ldots, \{v_n\}\}$, then $q_G(\A) = - \sum (\deg_G(v)/\vol_G(V))^2 < 0$.

The maximum \emph{modularity} $q^*(G)$ is defined as the maximum of $q_G(\A)$ over all possible partitions $\A$ of $V$; that is, $q^*(G) = \max_{\A} q_G(\A).$ In order to maximize $q_G(\A)$ one wants to find a partition with large edge contribution subject to small degree tax. If $q^*(G)$ approaches 1 (which is the trivial upper bound), we observe a strong community structure; conversely, if $q^*(G)$ is close to zero (which is the trivial lower bound), there is no community structure. The definition in~(\ref{eq:q_G_A}) can be generalized to weighted edges (with weight function $w: E \to \R_+$), by replacing edge counts with sums of the corresponding edge weights.

%%%%%%%%%%%%%%%%%%%%%%%%%%%%%%%%%%%%%%%%%%%%%%%%%%%%%%%%%%%
\subsubsection*{Using Graph Modularity for Hypergraphs}

%% FCT - check 2-section weights

Given a hypergraph $H=(V,E)$, it is common to transform its hyperedges into complete graphs (cliques), the process known as forming the 2-section of $H$ or clique expansion, the graph $H_{[2]}$, on the same set of nodes as $H$. For each hyperedge $e \in E$ with $|e| \ge 2$ having weight $w(e)$, 
$\binom{|e|}{2}$ 
edges are formed, each of them with weight of $w(e)/\binom{|e|}{2}$.
This choice preserves the total weight. There are other natural choices for the weight, for example the weighting scheme where $w(e)/(|e|-1)$ that ensures that the degree distribution of the created graph matches the one of the original hypergraph $H$~\cite{Kumar1,Kumar2}.
% Moreover, let us also mention that it also nicely translates a natural random walk on $H$ into a random walk on the corresponding $H_{[2]}$. % ~\cite{Francois-School}. 
As hyperedges in $H$ usually overlap, this process creates a multigraph. In order for $H_{[2]}$ to be a simple graph, if the same pair of vertices appear in multiple hyperedges, the corresponding edge weights are summed.

One of the approaches for finding communities in hypergraphs that practitioners use is to apply one of the algorithms that aim to maximize the original, graph modularity function (such as \textbf{Louvain}, \textbf{Leiden}, or \textbf{ECG}) to graph $H_{[2]}$. Despite the fact that this procedure is simple, it has a drawback that the 2-section graph looses some potentially useful information. Therefore, it is desired to define modularity function that is tailored explicitly for hypergraphs and aim to optimize it directly.

%%%%%%%%%%%%%%%%%%%%%%%%%%%%%%%%%%%%%%%%%%%%%%%%%%%%%%%%%%%
\subsubsection*{Hypergraph Modularity}

For edges of size greater than 2, several definitions can be used to quantify the edge contribution for a given partition $\A$ of the set of nodes. As a result, the choice of hypergraph modularity function is not unique. It depends on how strongly one believes that a hyperedge is an indicator that some of its vertices fall into one community. The fraction of nodes of a given hyperedge that belong to one community is called its \emph{homogeneity} (provided it is more than 50\%). In one extreme case, all vertices of a hyperedge have to belong to one of the parts in order to contribute to the modularity function; this is the \emph{strict} variant assuming that only homogeneous hyperedges provide information about underlying community structure. In the other natural extreme variant, the \emph{majority} one, one assumes that edges are not necessarily homogeneous and so a hyperedge contributes to one of the parts if more than 50\% of its vertices belong to it; in this case being over 50\% is the only information that is considered relevant for community detection. All variants in between guarantee that hyperedges contribute to at most one part. This is an important difference from the modularity on $H_{[2]}$, where a single original hyperedge is split into multiple graph edges that could be considered as contributing to multiple different parts (communities).
Once the variant is fixed, one needs to benchmark the corresponding edge contribution using the degree tax computed for the generalization of the Chung-Lu model to hypergraphs proposed in~\cite{kaminski2019clustering}.

The hypergraph modularity function is controlled by \emph{hyper-parameters} $\eta_{c,d} \in [0,1]$ ($d \ge 2$, $\lfloor d/2 \rfloor + 1 \le c \le d$). 
For a fixed set of hyper-parameters and a given partition $\A = \{A_1, A_2, \ldots, A_k\}$ of $V$, we define
\begin{equation}\label{eq:new_modularity}
q_{H}({\mathbf A}) = \sum_{d \geq 2} ~ \sum_{c = \lfloor d/2 \rfloor + 1}^{d} \eta_{c,d} \ q_{H}^{c,d}({\mathbf A}),
\end{equation}
where
$$
q_{H}^{c,d}({\mathbf A}) = \frac{1}{|E|} \sum_{A_i \in {\bf A}} 
\left( e_H^{c,d}(A_i) - |E_d| \cdot \Pr \left( \textrm{Bin} \left( d, \frac{\vol(A_i)}{\vol(V)} \right) = c \right) \right);
$$
$e_H^{c,d}(A_i)$ is the number of hyperedges of size $d$ that have exactly $c$ members in $A_i$, and $\textrm{Bin} (d,p)$ is the binomial random variable, that is,
$$
\Pr \left( \textrm{Bin} \big( d, p \big) = c \right) = \binom{d}{c} p^c (1-p)^{d-c}.
$$

Hyper-parameters $\eta_{c,d}$ give us a lot of flexibility and allow to value some hyperedges more than other ones depending on their size and homogeneity. However, there is a natural family of hyper-parameters that one might consider, namely, $\eta_{c,d} = (c/d)^{\tau}$ for some constant $\tau \in [0, \infty)$. We will refer to the corresponding modularity function as \emph{$\tau$-modularity function}. This family has only one parameter to tune, $\tau$, but it still covers a wide range of possible scenarios. For example, one might want to value all hyperedges equally ($\tau=0$) or value more homogeneous hyperedges more ($\tau > 0$), including the extreme situation in which only fully homogeneous hyperedges are counted ($\tau \to \infty$). In particular, we get the following four natural parameterizations of the modularity function to optimize: 
\begin{itemize}
\item \emph{strict modularity} ($\tau \to \infty$): $\eta_{d,d}=1$ and $\eta_{c,d}=0$ for $\lfloor d/2 \rfloor + 1 \le c < d$, 
\item \emph{quadratic modularity} ($\tau=2$): $\eta_{c,d}= (c/d)^2$ for $\lfloor d/2 \rfloor + 1 \le c \le d$,
\item \emph{linear modularity} ($\tau=1$): $\eta_{c,d}=c / d$ for $\lfloor d/2 \rfloor + 1 \le c \le d$,
\item \emph{majority modularity} ($\tau=0$): $\eta_{c,d}=1$ for $\lfloor d/2 \rfloor + 1 \le c \le d$.
\end{itemize}
Note that regardless of the parameter $\tau$, the weights are normalized so that $\max_c \eta_{c,d} = 1$ for all $d$. This ensures that the modularity function is normalized to be between 0 and 1. 

As already mentioned above, the choice of the parameter $\tau$ should be made depending on how much more homogeneous hyperedges are valued compared to inhomogeneous ones. However, in an absence of any external intuition about the nature of the ground-truth communities, our suggestion is to use $\tau = 2$. This choice is justified based on the connection to $H_{[2]}$, the corresponding 2-section graph of $H$. Indeed, hyperedges of size $d$ in $H$ that have exactly $c$ members in one of the communities contribute $\binom{c}{2} / \binom{d}{2} = \frac {c(c-1)}{d(d-1)} \approx (c/d)^2$ fraction of their original weight to the graph modularity function of $H_{[2]}$.

Having said that, let us stress the fact that optimizing the hypergraph $2$-modularity function of $H$ is \emph{not} equivalent to optimizing the graph modularity function of $H_{[2]}$ since hyperedges with $c \le d/2$ members in one of the communities do not contribute to the hypergraph modularity whereas they still do in the graph counterpart.

Indeed, this observation highlights the key difference between our approach to extracting communities from the hypergraph and doing it via the corresponding 2-section graph $H_{[2]}$ that we already indicated when introducing hypergraph modularity. Our assumption is that there exists an underlying set of latent communities in the hypergraph (commonly referred to as the ground-truth). A given set of nodes appears as a hyperedge with the probability that depends on whether the majority of them are from one of the communities or not. As a result, hyperedges of size $d$ in $H$ that have at most $d/2$ members in one of the communities are considered as noise, that unnecessarily influences the modularity function of $H_{[2]}$. Indeed, the hypergraph modularity is guaranteed to count a single hyperedge at most for one community (as we require $c>d/2$). On the other hand, the graph modularity of $H_{[2]}$ potentially treats a single hyperedge as a positive signal contributing to multiple communities.

\medskip

It is well known that optimizing modularity function in large networks might fail to resolve small communities, even when they are well defined. This well-known potential problem of applying a global null-models and is often referred to as the resolution limit~\cite{fortunato2007resolution}. A standard approach which tries to solve the resolution limit is to multiply the degree tax in the definition of the modularity function by a parameter $\gamma > 0$. This additional parameter controls the relative importance between the edge contribution and the degree tax. The hypergraph modularity function may be tuned the same way, if needed.

Finally, let us mention that for a given partition $\A$, the values of different modularity functions should not be compared, as they are scaled differently; rather the same modularity function should be used to rank various partitions for a given graph.

%\pc{Not sure if this is useful unless we want to say that comment about linear modularity being close to the 2-section. But does that support it?} Moreover,~(\ref{eq:new_modularity}) may well approximate the graph modularity for the corresponding 2-section graph $H_{[2]}$. Indeed, if $c$ vertices of a hyperedge $e$ of size $d$ and weight $w(e)$ fall into one part of the partition $\A$, then the contribution to the graph modularity is $w(e) {c \choose 2} / (|e|-1)$ (in the variant where the degrees are preserved) or $w(e) {c \choose 2} / {|e| \choose 2} \approx w(e) (c/|e|)^2$ (if the total weight is preserved).
%Hence, the hyper-parameters can be adjusted to reflect that. The only difference is that~(\ref{eq:new_modularity}) does not allow to include contributions from parts that contain at most $d/2$ vertices which still contributes to the graph modularity of $H_{[2]}$. However, most of the contribution comes from large values of $c$ and so the two corresponding measures are close in practice.

%%%%%%%%%%%%%%%%%%%%%%%%%%%%%%%%%%%%%%%%%%%%%%%%%%%%%%%%%%%
\section{Hypergraphs Used in Our Experiments}\label{sec:hypergraphs}
%%%%%%%%%%%%%%%%%%%%%%%%%%%%%%%%%%%%%%%%%%%%%%%%%%%%%%%%%%%

In this section, we introduce the hypergraphs we use in our experiments, both synthetic and real-world ones.

%%%%%%%%%%%%%%%%%%%%%%%%%%%%%%%%%%%%%%%%%%%%%%%%%%%%%%%%%%%
\subsection{Synthetic Hypergraph Model: \textbf{h--ABCD}}

There are very few hypergraph datasets with ground-truth identified and labelled. Synthetic networks are extremely useful to test various scenarios, such as the level of noise, via tuneable and interpretable parameters. As a result, there is need for synthetic random graph models with community structure that resemble real-world networks in order to benchmark and tune clustering algorithms that are unsupervised by nature. 

It is worth mentioning that the family of clustering algorithms we are interested in aims to find partitions that maximize given modularity function, not to find the ground-truth partition. Those are often very similar partitions (but not always). Note that ground truth partitions typically influence the creation of a hypergraph in a noisy way, which means that just as a consequence of this randomness a good community in a graph (after the randomness is resolved) does not have to match exactly the ground truth community.

In particular, algorithm $A$ would be considered better than algorithm $B$ if it finds a partition yielding larger modularity. Selecting the right modularity function to optimize is crucial for making sure that the outcome of the algorithm is close to the ground-truth (or some specific requirements of the user), but once the function is selected the algorithm should aim to maximize it. We propose a simple, unsupervised method for making such selection (see Section~\ref{sec:modularity_selection}) but this paper focuses on the optimization algorithm.

The standard for the generation of synthetic graphs is rather clear. The \textbf{LFR} (\textbf{L}ancichinetti, \textbf{F}ortunato, \textbf{R}adicchi) model~\cite{lancichinetti2008benchmark,lancichinetti2009benchmarks} generates networks with communities and at the same time it allows for the heterogeneity in the distributions of both node degrees and of community sizes. It became a standard and extensively used method for generating artificial networks. The \textbf{A}rtificial \textbf{B}enchmark for \textbf{C}ommunity \textbf{D}etection (\ABCD)~\cite{kaminski2021artificial} was recently introduced and implemented\footnote{\url{https://github.com/bkamins/ABCDGraphGenerator.jl/}}, including a fast implementation\footnote{\url{https://github.com/tolcz/ABCDeGraphGenerator.jl/}} that uses multiple threads (\textbf{ABCDe})~\cite{kaminski2022abcde}. Undirected variant of \textbf{LFR} and \textbf{ABCD} produce graphs with comparable properties but \textbf{ABCD}/\textbf{ABCDe} is faster than \textbf{LFR} and can be easily tuned to allow the user to make a smooth transition between the two extremes: pure (disjoint) communities and random graph with no community structure. Moreover, it is easier to analyze theoretically---for example, in~\cite{kaminski2022modularity,barrett2023abcd} various theoretical asymptotic properties of the \ABCD\ model are investigated including the modularity function and self-similarities of the ground-truth communities.

The situation for hypergraphs is not as clear as for graphs. There are not only few real-world datasets (with ground-truth) available, but also there are not so many synthetic hypergraph models. Fortunately, the building blocks in the \ABCD\ model are flexible and may be adjusted to satisfy different needs. For example, the model was adjusted to include potential outliers in~\cite{kaminski2023artificial} resulting in \textbf{ABCD+o} model. Adjusting the model to hypergraphs is more complex but it was also done recently~\cite{kaminski2023hypergraph} resulting in \textbf{h--ABCD} model. We will use this model for our experiments. 

The \textbf{h--ABCD} model generates a hypergraph on $n$ nodes. The degree distribution follows power-law with exponent $\gamma$, minimum and maximum value equal to $\delta$ and, respectively, $D$. Community sizes are between $s$ and $S$, and also follow power-law distribution, but this time with exponent $\beta$. 
% The suggested range of values for parameters $\gamma$ and $\beta$ are chosen according to experimental values commonly observed in complex networks not only represented as graphs~\cite{barabasi2016network,orman2009comparison} but also as hypergraphs~\cite{do2020structural}. 
Parameter $\xi$ is responsible for the level of noise. If $\xi = 0$, then each hyperedge is a community hyperedge meaning that majority of its nodes belong to one community. On the other extreme, if $\xi=1$, then communities do not play any roles and hyperedges are simply ``sprinkled'' across the entire hypergraph that we will refer to as background hypergraph. Vector $(q_1, \ldots, q_L)$ determines the distribution of the number of hyperedges of a given size, where $L$ is the size of largest hyperedges. 

Finally, parameters $w_{c,d}$ specify how many nodes from its own community a given community hyperedge should have. We call a community hyperedge to be of type $(c,d)$ if it has size $d$ and exactly $c$ of its nodes belong to one of the communities. Note that, in light of the discussion we had at the end of the previous section, we require that a community hyperedge must have more than a half of its nodes from the community. Therefore, $w_{c,d}$ is defined for $d/2<c\leq d$, where $d\in[L]$.

The model is flexible and may accept any family of parameters $w_{c,d}$ satisfying specific needs of the users, but here is a list of three standard options implemented in the code: 
\begin{itemize}
    \item {\em majority} model: $w_{c,d}$ is uniform for all admissible values of $c$, that is, for any $d/2<c\leq d$,
    $w_{c,d} = \frac {1}{(d - \lfloor d/2 \rfloor)} = \frac {1}{\lceil d/2 \rceil},$
    \item {\em linear} model: $w_{c,d}$ is proportional to $c$ for all admissible values of $c$, that is, for any $d/2<c\leq d$,
    $w_{c,d} = \frac {2c}{(d+\lfloor d/2 \rfloor + 1)(d - \lfloor d/2 \rfloor)} = \frac {2c}{(d+\lfloor d/2 \rfloor + 1) \lceil d/2 \rceil},$
    \item {\em strict} model: only ``pure'' hyperedges are allowed, that is
    $w_{d,d}=1$ and $w_{c,d}=0$ for $d/2 < c < d$.
\end{itemize}
Let us note that the parameterizations of $w_{c,d}$ in \textbf{h--ABCD} and $\eta_{c,d}$ in the definition of the hypergraph modularity function have the same (due to matching functional form) name but they are not equivalent. Parameters $w_{c,d}$ determine the composition of hyperedges in the generated synthetic graph whereas parameters $\eta_{c,d}$ specify the objective function that the analyst decided to optimize against while looking for communities in the hypergraph at hand.

\medskip

Specifically, we used the following parameters for our experiments:
\begin{itemize}
    \item $n = 300$ nodes,
    \item power-law degree exponent $\alpha = 2.5$, in the range $[5,30]$,
    \item power-law community size exponent $\beta = 1.5$, in the range $[80,120]$. % \pc{It forces that we have exactly 3 communities. So it is weird to call it "power-law distribution"? ;-) Should we comment on it or just leave it as is?}
\end{itemize}
We generated 6 families of \textbf{h--ABCD} hypergraphs, namely:
\begin{itemize}
    \item {\tt linear\_2to5}: linear model for $w_{c,d}$, with edge sizes 2 to 5 (with respective probabilities $0.1, 0.4, 0.4, 0.1$),
    \item {\tt majority\_2to5}: majority model for $w_{c,d}$, with edge sizes 2 to 5 (with respective probabilities $0.1, 0.4, 0.4, 0.1$),
    \item {\tt strict\_2to5}: strict model for $w_{c,d}$, with edge sizes 2 to 5 (with respective probabilities $0.1, 0.4, 0.4, 0.1$),
    \item {\tt linear\_5}: linear model for $w_{c,d}$, with all edge of size 5,
    \item {\tt majority\_5}: majority model for $w_{c,d}$, with all edge of size 5, and  
    \item {\tt strict\_5}: strict model for $w_{c,d}$, with all edge of size 5.
\end{itemize}

%%%%%%%%%%%%%%%%%%%%%%%%%%%%%%%%%%%%%%%%%%%%%%%%%%%%%%%%%%%
\subsection{Real-world Hypergraphs}
\label{sec:real_hypergraphs}

To illustrate various aspects of hypergraph modularity-based clustering, we analyze a few real-world hypergraphs. The first is a contact hypergraphs in which nodes correspond to primary school children or teachers and hyperedges represent close physical proximity between individuals within a prescribed time period (see~\cite{chodrow2021generative,stehle2011,mastrandrea2015} for more details). 
There are 242 nodes labelled with respect to their class (there are 10 classes), plus another label for the teachers. There are 12,704 hyperedges of size up to 5. 
% In the second network ({\tt high-school}), nodes correspond to high school students. This time, there are 327 nodes labelled with respect to their class (there are 9 classes), and there are 7,818 edges of size up to 5. 
In Table~\ref{tab:edges_real} (left), we show the distribution of edge composition for this dataset with respect to the ground-true communities. We see that there are many edges between communities, in particular edges of size 2. Community edges (with $c>d/2$) are mostly ``pure'' edges (with $c=d$), but there is a significant number of edges of type $(c,d)=(2,3)$, thus it is unclear if hypergraph $\tau$-modularity functions with large parameter $\tau$ would do well, or if a small value for $\tau$ should be used.

\begin{table}[ht]
\centering
\begin{tabular}{ccrc}
\multicolumn{4}{c}{\tt primary-school} \\

\toprule
d & c & purity & frequency \\
\midrule
2 & 1 & 50\% & 5202 \\
\rowcolor{lightgray}
2 & 2 & 100\% & 2546  \\
\rowcolor{lightgray}
3 & 3 & 100\% & 2434 \\
\rowcolor{lightgray}
3 & 2 & 67\% & 1751 \\
3 & 1 & 33\% & 415 \\
\rowcolor{lightgray}
4 & 4 & 100\% & 158  \\
4 & 2 & 50\% & 93 \\
\rowcolor{lightgray}
4 & 3 & 75\% & 84 \\
4 & 1 & 25\% & 12 \\
\rowcolor{lightgray}
5 & 3 & 60\% & 6 \\
% 5 & 1 & 20\% & 0 \\
% 5 & 2 & 40\% & 1 \\
% \rowcolor{lightgray}
% 5 & 4 & 80\% & 1 \\
% \rowcolor{lightgray}
% 5 & 5 & 100\% & 1 \\
\bottomrule
\end{tabular}
\hspace{1.0cm}
\begin{tabular}{ccrc}
\multicolumn{4}{c}{\tt cora} \\
\toprule
d & c & purity & frequency \\
\midrule
\rowcolor{lightgray}
2 & 2 & 100\% & 472 \\
\rowcolor{lightgray}
3 & 3 & 100\% & 307 \\
\rowcolor{lightgray}
4 & 4 & 100\% & 175 \\
2 & 1 & 50\% & 151 \\
\rowcolor{lightgray}
3 & 2 & 67\% & 118 \\
\rowcolor{lightgray}
4 & 3 & 75\% & 91 \\
\rowcolor{lightgray}
5 & 5 & 100\% & 83 \\
\rowcolor{lightgray}
5 & 4 & 80\% & 55 \\
4 & 2 & 50\% & 42 \\
3 & 1 & 33\% & 39 \\
% 5 & 3 & 34 \\
% 5 & 2 & 7 \\
% 4 & 1 & 4 \\
% 5 & 1 & 1 \\
\bottomrule
\end{tabular}
\vspace{.25cm}
\caption{The number of hyperedges of type $(c,d)$ (the top-10 most frequent ones) for the \texttt{primary-school} dataset (left) and the \texttt{cora} co-citation dataset (right).
(Combinations contributing to hypergraph modularity are highlighted in grey.)}
\label{tab:edges_real}
\end{table}

Another dataset we use for our experiments in the co-reference dataset between scientific publication which belong to one of seven classes ({\tt cora}); see~\cite{yadati2019} for more details. Hyperedges consist of co-cited scientific publications, and we only keep hyperedges of size 2 or more. There are 1,434 nodes appearing in at least one hyperedge (cited publications), and 1,579 hyperedges.
In Table~\ref{tab:edges_real} (right), we show the top-10 distribution of edge composition with respect to the true communities. 
We see that there are many pure community edges (with $c = d$), so we can expect that hypergraph $\tau$-modularity functions with large parameter $\tau$ would do well.

%%%%%%%%%%%%%%%%%%%%%%%%%%%%%%%%%%%%%%%%%%%%%%%%%%%%%%%%%%%
\section{Hypergraph Modularity Optimization Algorithm: \textbf{h--Louvain}}
\label{sec:algorithm}
%%%%%%%%%%%%%%%%%%%%%%%%%%%%%%%%%%%%%%%%%%%%%%%%%%%%%%%%%%%

Let us fix the hypergraph modularity function $q_H(\A)$, either by restricting ourselves to $\tau$-modularity function with some specific value of $\tau$ (such as $\tau=2$ that is recommended as the default value) or by specifying the more general hyper-parameters $\eta_{c,d}$. The goal of this section is to highlight challenges in designing a heuristic algorithm aiming to optimize $q_H(\A)$ and to describe our solution that overcame these challenges, producing an algorithm that we will refer to as \textbf{h--Louvain}.

%%%%%%%%%%%%%%%%%%%%%%%%%%%%%%%%%%%%%%%%%%%%%%%%%%%%%%%%%%%
\subsection{\textbf{Louvain} Algorithm}\label{ssec:Louvain}

Let us start by introducing one of the most popular algorithms for detecting communities in graphs, namely, the \textbf{Louvain} algorithm~\cite{blondel2008fast}. It is a hierarchical clustering algorithm that tries to optimize the modularity function we described in Section~\ref{sec:modularity}. 

In the first pass of this algorithm, small communities are found by optimizing the graph modularity function locally on all nodes. Then, each small community is grouped together into a single node that we will refer to as super-node. This process is repeated recursively on those smaller graphs consisting of super-nodes (the subsequent passes) until no improvement on the modularity function can be further achieved.

One pass of the algorithm consists of two phases that are repeated iteratively. In the first phase, each node in the network is assigned to its own community. For each node $v$, we consider all neighbours $u$ of $v$ and compute the change in the modularity function if $v$ is removed from its own community and moved into the community of $u$. It is important to mention that this value can be easily and efficiently calculated without the need to recompute the modularity function from scratch. Once all the communities that $v$ could belong to are considered, $v$ is placed into the community that resulted in the largest increase of the modularity function. If no increase is possible, $v$ remains in its original community. 
The process is repeated for the remaining nodes following a given (typically random) permutation of nodes, possibly multiple times, until a local maximum value is achieved and the first phase ends.

During the second phase, the algorithm contracts all nodes that belong to one community into a single super-node. All edges within that community are replaced by a single weighted loop. Similarly, all edges between two communities are replaced by a single weighted edge. Once the new network is created, the second phase ends. The resulting graph is typically much smaller than the original graph. As a result, the first pass is typically the most time consuming part of the algorithm.

%%%%%%%%%%%%%%%%%%%%%%%%%%%%%%%%%%%%%%%%%%%%%%%%%%%%%%%%%%%
\subsection{Challenges with Adjusting the Algorithm to Hypergraphs}\label{ssec:challenges}

One could try to directly apply the \textbf{Louvain} algorithm to optimize hypergraph modularity, since in both cases the goal is to find a partition of the set nodes. However, as the algorithm moves only one node at a time, it creates a problem in the case of hypergraphs.

Consider, for example, a hypergraph in which all hyperedges have size at least four. In this case, regardless which two nodes $u$ and $v$ are considered for possible merging into one community, the edge contribution would not change (that is, it would stay equal to zero), even if $u$ and $v$ are part of some hyperedge. (Recall that only hyperedges with majority of nodes from the same community may affect the edge contribution). On the other hand, the degree tax would increase after such a move and, as a result, the modularity function would decrease. Therefore, no move would be made and the algorithm would get immediately stuck. We will refer to this issue as a \emph{lift off from the ground} problem.

The above, extreme, situation is not the only problem one should be aware of. This time consider a hypergraph that consists of a mixture of hyperedges of various sizes, including edges of size two. In this scenario there is no problem with lifting off from the ground but small hyperedges clearly play a much more important role than large ones during the initial merging in the first phase of the algorithm. On the other hand, very large hyperedges would be mostly ignored. This behaviour is not desirable either. In order to illustrate a potential danger, consider a hypergraph representing interactions between researchers at some institution. Nodes in this hypergraph correspond to researchers and hyperedges correspond to meetings of some groups of people. For simplicity, assume that there are two communities, say, faculty of science and faculty of engineering. Many hyperedges within the two communities are large (e.g.\ hyperedges associated with departmental meetings) whereas hyperedges between the two communities are mostly of size two (e.g.\ two members of different teams meet individually from time to time). In this scenario, the algorithm would start merging people from different communities during the first phase. 

Finally, let us note that one could alternatively consider modifying the algorithm and allow for not only merging two nodes into one community in a single move but entire hyperedges. Again, this does not seem to be desirable as hyperedges might consist of members from different communities and so such operations would generate many incorrect merges too fast.

%%%%%%%%%%%%%%%%%%%%%%%%%%%%%%%%%%%%%%%%%%%%%%%%%%%%%%%%%%%
\subsection{Our Approach to Hypergraph Modularity Optimization}
\label{subsec:algorithm}

In order to overcome the above mentioned challenges, we want to design an algorithm that, as in the classical \textbf{Louvain} algorithm, merges single pairs of nodes while, at the same time, takes into account information stored in hyperedges of all sizes. To that end we propose to optimize a linear combination of the hypergraph modularity $q_{H}({\mathbf A})$ and the graph modularity of the corresponding 2-section graph $H_{[2]}$, that is, optimize function
\begin{equation}
q ({\mathbf A}, \alpha) := \alpha \cdot q_{H}({\mathbf A}) + (1-\alpha) \cdot q_{H_{[2]}} ({\mathbf A}),
\label{eq:optimize}
\end{equation}
where $\alpha \in [0,1]$. For simplicity, we will refer to our algorithm as \textbf{h--Louvain}.

To understand the motivation behind this approach, let us observe the following. The hypergraph modularity, equation~(\ref{eq:new_modularity}), is flexible and may approximate well the graph modularity for the corresponding 2-section graph $H_{[2]}$. Indeed, if $c$ vertices of a hyperedge $e$ of size $d$ and weight $w(e)$ fall into one part of the partition $\A$, then the contribution to the graph modularity is $w(e) \binom{c}{2} / \binom{|e|}{2} \approx w(e) (c/|e|)^2$ (in the variant of the 2-section where the total weight is preserved) or $w(e) \binom{c}{2} / (|e|-1)$ (if the degrees are preserved). Hence, the hyper-parameters of the hypergraph modularity can be adjusted to approximate $H_{[2]}$ modularity. The only difference is that~(\ref{eq:new_modularity}) does not allow to include contributions from parts that contain at most $d/2$ vertices which still contributes to the graph modularity of $H_{[2]}$.

The observation justifies using $q ({\mathbf A}, \alpha)$ for optimizing the hypergraph modularity. It is a linear combination of the actual hypergraph modularity we want to optimize, $q_H({\mathbf A})$, and an approximation of the hypergraph modularity for special value of hyper-parameter ($\tau=2$) and without the restriction of hyperedge contribution, $q_{H_{[2]}} ({\mathbf A})$. The benefit of the second part is that it is sensitive to merging two nodes and so it always gives some indication of how nodes should be merged (even if the first part $q_H(\mathbf A)$ does not give such an indication). In short, it resolves the \emph{lifting off from the ground} problem. If $\alpha$ is close to zero, then we concentrate mostly on the approximation part, while if $\alpha$ is close to one, then we mostly concentrate on the actual hypergraph modularity we aim to optimize.

The above discussion leads us to the conclusion that the parameter $\alpha \in [0,1]$ should be appropriately tuned during the algorithm. The main questions are: a) when the change should be made, and b) what values of this parameter should be used? In~\cite{kaminski2023modularity}, we performed various experiments and made the following observations. The optimization process should start with low values of the parameter $\alpha$ (to let the process \emph{lift off from the ground}) and then it should be gradually increased till it reaches one by the end of the process. The algorithm should start increasing parameter $\alpha$ when the communities induce enough edges so that merging additional nodes makes a difference in the edge contribution of the $q_H$ function value; this, in particular, means that since the strict hypergraph modularity pays attention to only pure hyperedges (all members belong to one community), in this case, the algorithm needs to start with lower values of $\alpha$ and increase it slower than for the majority or the linear counterparts of the hypergraph modularity for which it is enough that over 50\% of nodes in some hyperedge are captured in one community.

Based on these observations, we propose the following schema for setting the successive values of $\alpha$ used in the objective function~(\ref{eq:optimize}), which leads to monotonic (non-decreasing) sequences ($\alpha_1,\alpha_2,\dots$). The schema is guided by the following two parameters: $p_b \in [0,1]$ and $p_c \in (0,1)$. The parameter $p_b$ is used to determine the values of $\alpha_i$, while $p_c$ governs when the algorithm switches from $\alpha_{i-1}$ to $\alpha_{i}$ (for $i \ge 2$) as the optimization progresses.

For a given pair of parameters $(p_b, p_c)$, the values of $\alpha_i$ are determined as follows: for any $i \in \N$,
$$
\alpha_i = 1 - (1 - p_b)^{i-1}.
$$
(We use the convention that $0^0=1$.) Note that $\alpha_1=0$ and $\alpha_i \to 1$ as $i \to \infty$, unless $p_b=0$. In the degenerate case, if $p_b=0$, then $\alpha_i=0$ for all $i$.
The algorithm switches from $\alpha_{i-1}$ to $\alpha_i$'s (for $i \ge 2$) when the number of communities drops to $np_c^{i-1}$ or below for the first time (note that the number of communities typically decreases but it is not always the case; as usual, $n$ denotes the number of nodes).
In summary, the two parameters have the following interpretation: $p_b$ controls the rate of change of $\alpha$ (values close to zero make $\alpha_i$ converge to one slowly, values close to one make convergence fast); $p_c$ controls the speed of change of $\alpha$.

There are two possible endings once the algorithm reaches a partition in which no improvement of the modularity function is possible via local changes. (Note that it might happen when the value of $\alpha_i$ is still away from one.) By default, we fix $\alpha_i=1$ and continue optimizing the hypergraph modularity function on the small graph consisting of super-nodes until no further improvement can be achieved. Alternatively, the local optimization can be performed on the original graph consisting of nodes. The pseudo-code of \textbf{h--Louvain} can be found in the Appendix (see Section~\ref{apendix_code}). As it is discussed in Section~\ref{sec:bayesopt}, we use the default ending when doing Bayesian optimization to select a good pair $(p_b, p_c)$ of parameters because it is faster. Once the final pair is chosen, we do an additional tuning process with local-optimization enabled which typically yields better values of the objective function. 

\subsection{Parameters of the h--Louvain Algorithm}\label{ssec:parameters}

In this subsection, we aim to investigate the quality of the \textbf{h--Louvain} algorithm for different pairs of parameters $(p_b, p_c)$. To that end, we analyzed the performance of the algorithm using $9$ different \textbf{h--ABCD} graphs on $1{,}000$ nodes. For each of the three options for community hyperedges in the \textbf{h--ABCD} model (namely, strict, linear, and majority), we used the following three settings with respect of different levels of noise and sizes of hyperedges: 
\begin{enumerate}
\item small level of noise ($\xi = 0.15$, $\xi_{emp} = 0.29$), hyperedges of size between 2 and 5, the degree distribution following power-law with exponent $\gamma = 2.5$, minimum and maximum degree $5$ and $20$, 
\item large level of noise ($\xi = 0.6$, $\xi_{emp} = 0.62$), hyperedges of size between 2 and 5, the degree distribution following power-law with exponent $\gamma = 2.5$, minimum and maximum degree $5$ and $20$, 
\item large hyperedges (sizes between 5 and 8), large level of noise ($\xi = 0.3$, $\xi_{emp} = 0.63$), the degree distribution following power-law with exponent $\gamma = 2.5$, minimum and maximum degree $5$ and $60$. 
\end{enumerate}
(In the above description, $\xi_{emp}$ refers to the actual level of noise in the produced hypergraph. The model ensures that $\xi_{emp} \approx \xi$ for graphs without small communities. In our scenario, it is not the case but the generated hypergraphs still have drastically different levels of noise.)
In all three settings, the distribution of community sizes follows power-law with exponent $\beta = 1.5$, minimum and maximum size $10$ and $30$. The distribution of hyperedges of different sizes is $(0.4,0.3,0.2,0.1)$, that is, there are slightly more hyperedges of smaller size.

\begin{figure}
\centering
\includegraphics[width=0.32\textwidth]{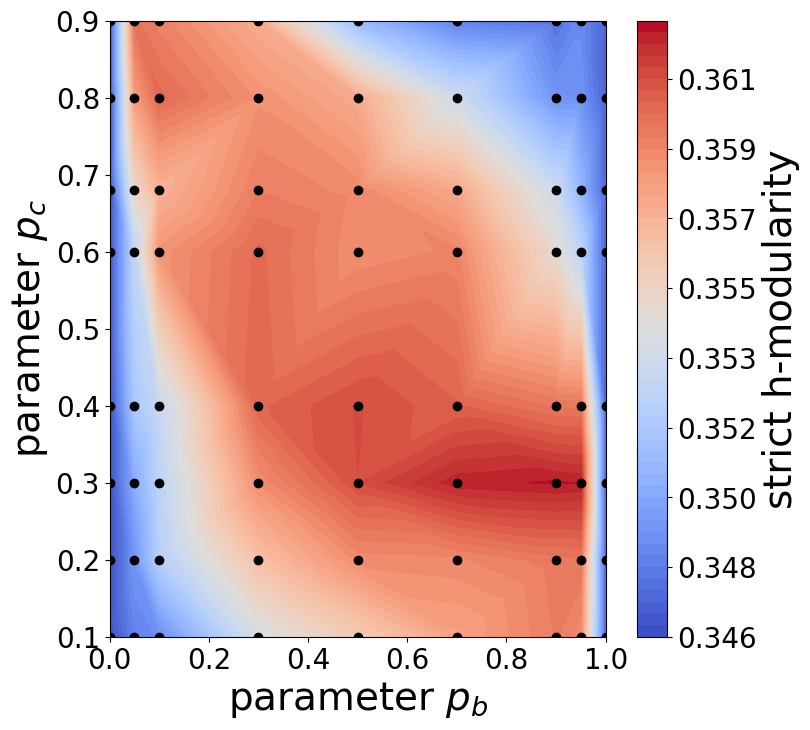}
\includegraphics[width=0.32\textwidth]{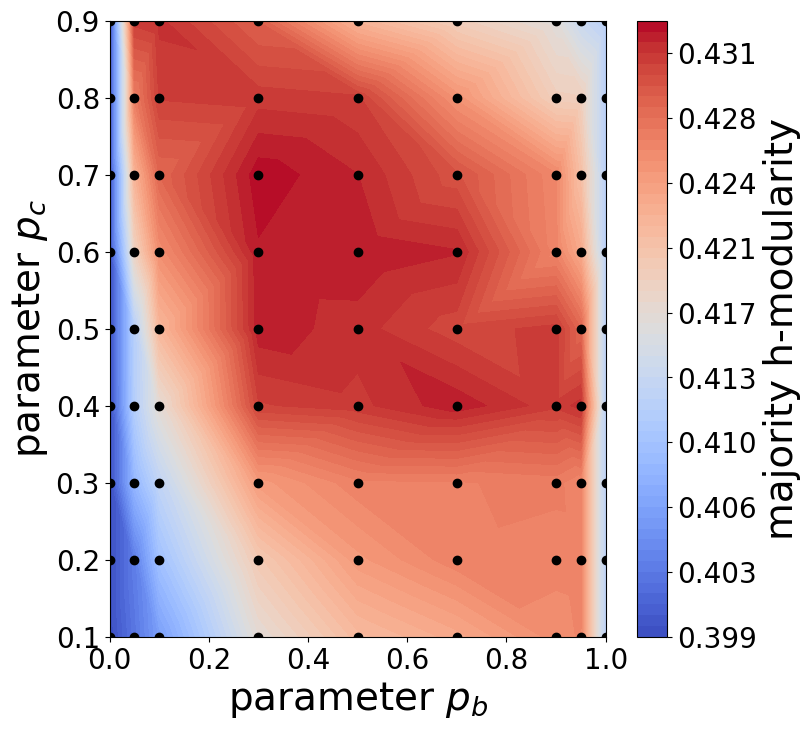}
\includegraphics[width=0.32\textwidth]{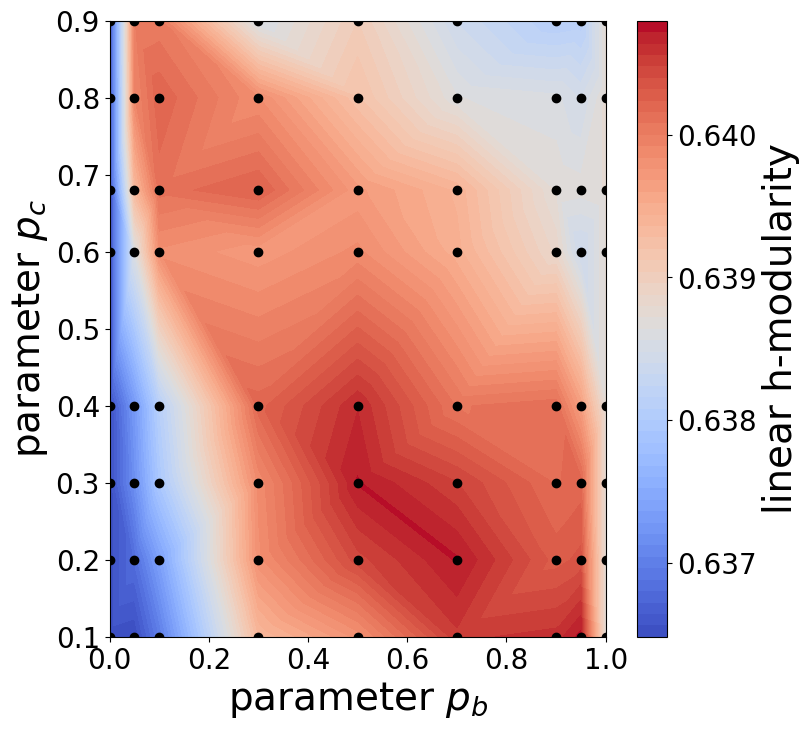}
\caption{Quality of \textbf{h--Louvain} on \textbf{h--ABCD} as a function of parameters $p_b, p_c$. Optimal combinations of the two parameters depend on the choice of \textbf{h--ABCD} variant and hypergraph modularity function: strict, large level of noise (left), majority, large level of noise (middle), or linear, small level of noise (right).}
\label{fig:heatmap-m-s}
\end{figure}

Figure~\ref{fig:heatmap-m-s} presents the performance of the algorithm for three selected hypergraphs out of the $9$ we experimented with. (Results for the remaining six hypergraphs can be found in the associated GitHub repository.) For each hypergraph, we present the quality of the algorithm optimizing the corresponding modularity function (that is, for example, for strict hypergraph we optimize the strict modularity function) as a function of $(p_b,p_c)$. The parameters were tested from the 2-dimensional grid $(p_b, p_c)$, where $p_b \in \{0.05,0.1,0.3,0.5,0.7,0.9,0.95\}$ and $p_c \in \{0.1, 0.2, \ldots, 0.9\}$. For each pair of parameters,  the average modularity function is reported over 10 independent runs with different random seeds. (Recall that \textbf{h--Louvain} is a randomized algorithm.)

The general conclusion is that the optimal choice of parameters depends on many factors: property of the hypergraph (such as the composition of community hyperedges, the level of noise, sizes of hyperedges) as well as the modularity function that one aims to maximize. However, not surprisingly, it is not recommended to set both parameters to be close to zero (the case of slow and small increases of the alpha parameter, so optimizing mainly the graph modularity of the corresponding 2-section graph $H_{[2]}$), or to be close to one (the case of fast and significant changes of the alpha parameter, so optimizing the hypergraph modularity $q_{H}$ almost from the beginning of algorithm execution). The optimal values are often obtained for settings with balanced values of both parameters, namely, with $p_b + p_c \approx 1$. In order to find a ``sweet spot'' in an unsupervised way, we use Bayesian optimization that we discuss next.

\subsection{Bayesian Optimization: Selecting the Parameters}\label{sec:bayesopt}

In order to find a good pair of the two parameters $(p_b, p_c)$ guiding the \textbf{h--Louvain} algorithm that yield large hypergraph modularity function, we use the Bayesian optimization approach~\cite{frazier2018tutorial}. We chose this tool for our problem as this approach is best suited for optimizing objective functions that take a long time to evaluate over continuous domains of less than 20 dimensions, and tolerates well non-negligible local variability of the evaluation of the function. It builds a surrogate for the objective function and quantifies the uncertainty in that surrogate using a Bayesian machine learning technique, Gaussian process regression, and then uses an acquisition function defined from this surrogate to decide where to sample the domain in an on-line fashion.

Specifically, in our case the Bayesian optimization aims to explore the two dimensional space $(p_b, p_c)$ with $p_b \in [0,1]$ and $p_c \in (0,1)$. The target function is defined as the average modularity function of the outcome partition for $10$ independent executions of the \textbf{h--Louvain} algorithm with different (but fixed across runs) random number generator seeds. Note that in this setting we maximize a deterministic function (since the seeds are fixed). We take the average over 10 different seeds because we aim to identify the region of the $(p_b, p_c)$ domain that leads to good values of the obtained evaluations of modularity and taking the average reduces the noise that is present in modularity values observed in single runs of the algorithm.

Because tuning hyper-parameters is computationally intensive, we initially use the default ending of the algorithm, that is, without the local-optimization approach for the last phase (see Section~\ref{subsec:algorithm} for an explanation how the optional ending works). The reason for this choice is that in this phase of the process we are mostly interested in capturing the shape of the response surface (recall that we take the average of 10 runs of the algorithm for the very same reason, namely, to smooth-out the results and to better capture the shape of the response surface). The default ending is sufficient for this purpose and it is substantially faster. After finding an approximation of the optimal $(p_b, p_c)$ combination, the algorithm comes back to the partition obtained with these parameters but this time the local-optimization approach is used during the last phase. It is more computationally expensive, but at this stage of the procedure we are interested in finding the maximum value of the hypergraph modularity, and so this additional computational cost is justified.

We configured the Bayesian optimization procedure so that it starts with evaluation of $5$ initial pairs of parameters selected randomly from the domain and at least $10$ pairs are tested in total. Once the Bayesian optimization converges, the algorithm returns the partition of the largest modularity from all partitions generated during the entire process. Note that this partition might not be one of the 10 partitions that contributed to the largest value of the target function; these partitions only have the best average modularity.

\begin{figure*}[ht]
        \subfloat[Step 8]{%
            \includegraphics[width=.53\linewidth]{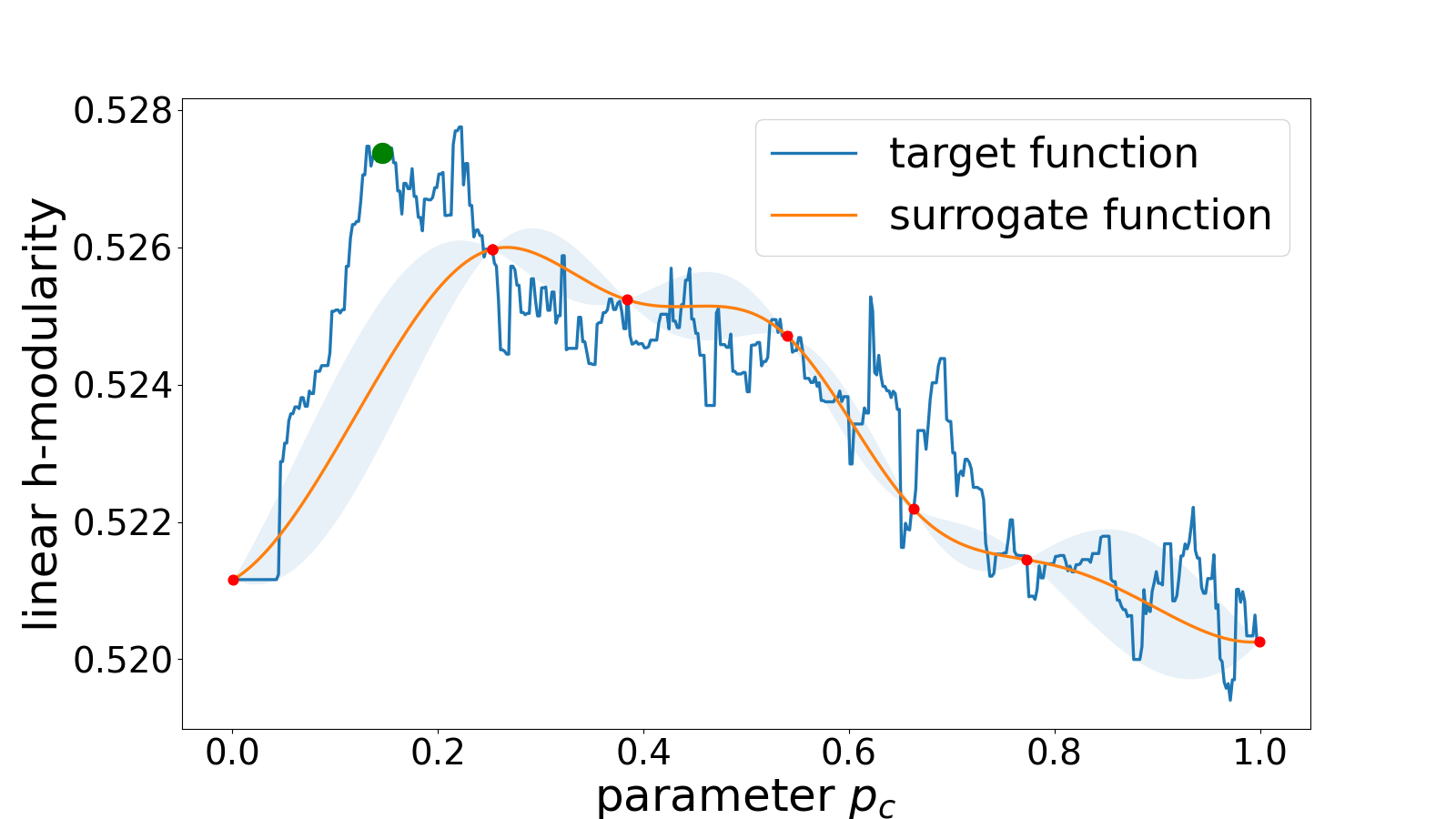}%
            \label{subfig:a}%
        }\hfill
        \subfloat[Step 9]{%
            \includegraphics[width=.53\linewidth]{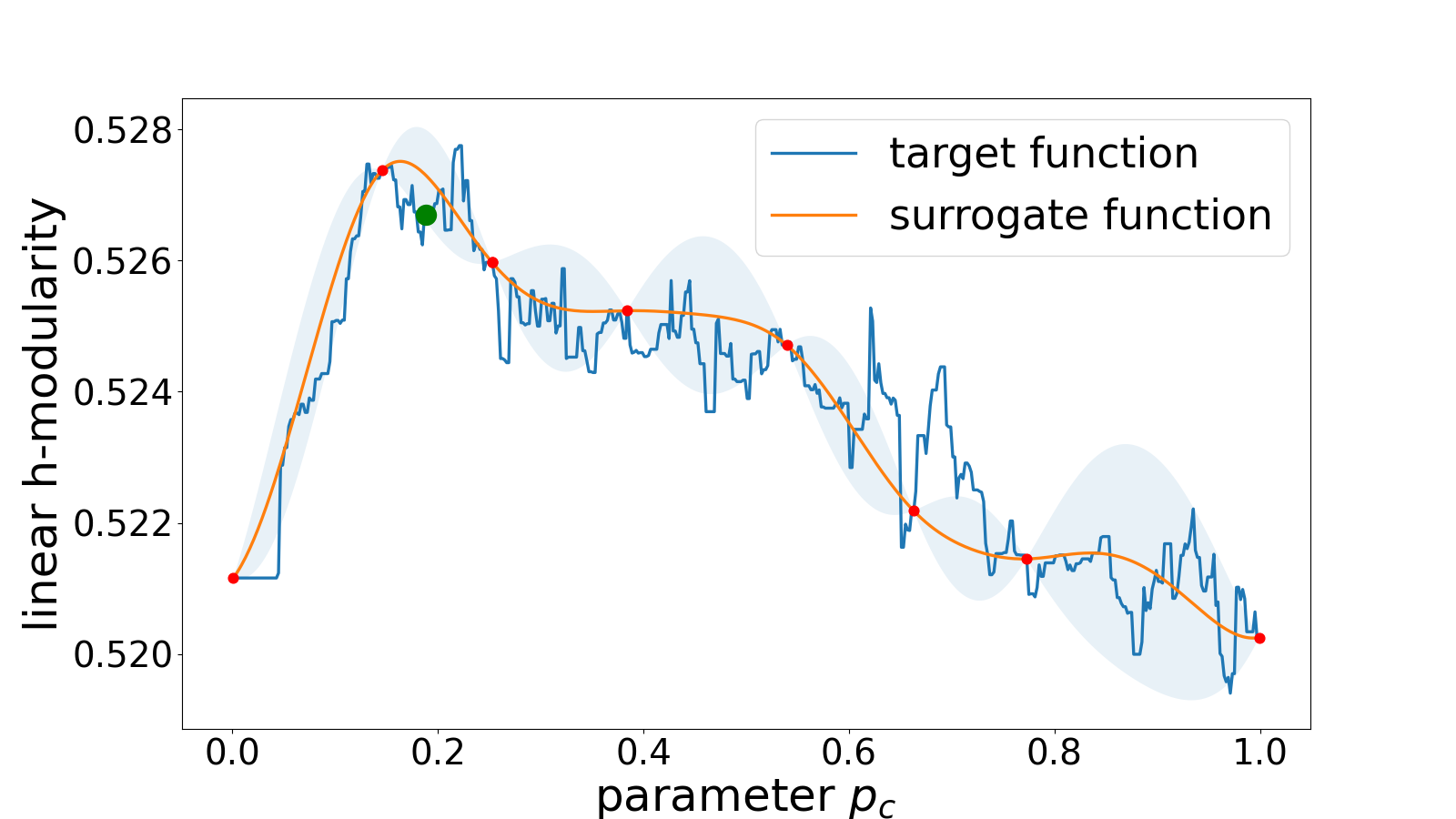}%
            \label{subfig:b}%
        }\\
        \subfloat[Step 10]{%
            \includegraphics[width=.53\linewidth]{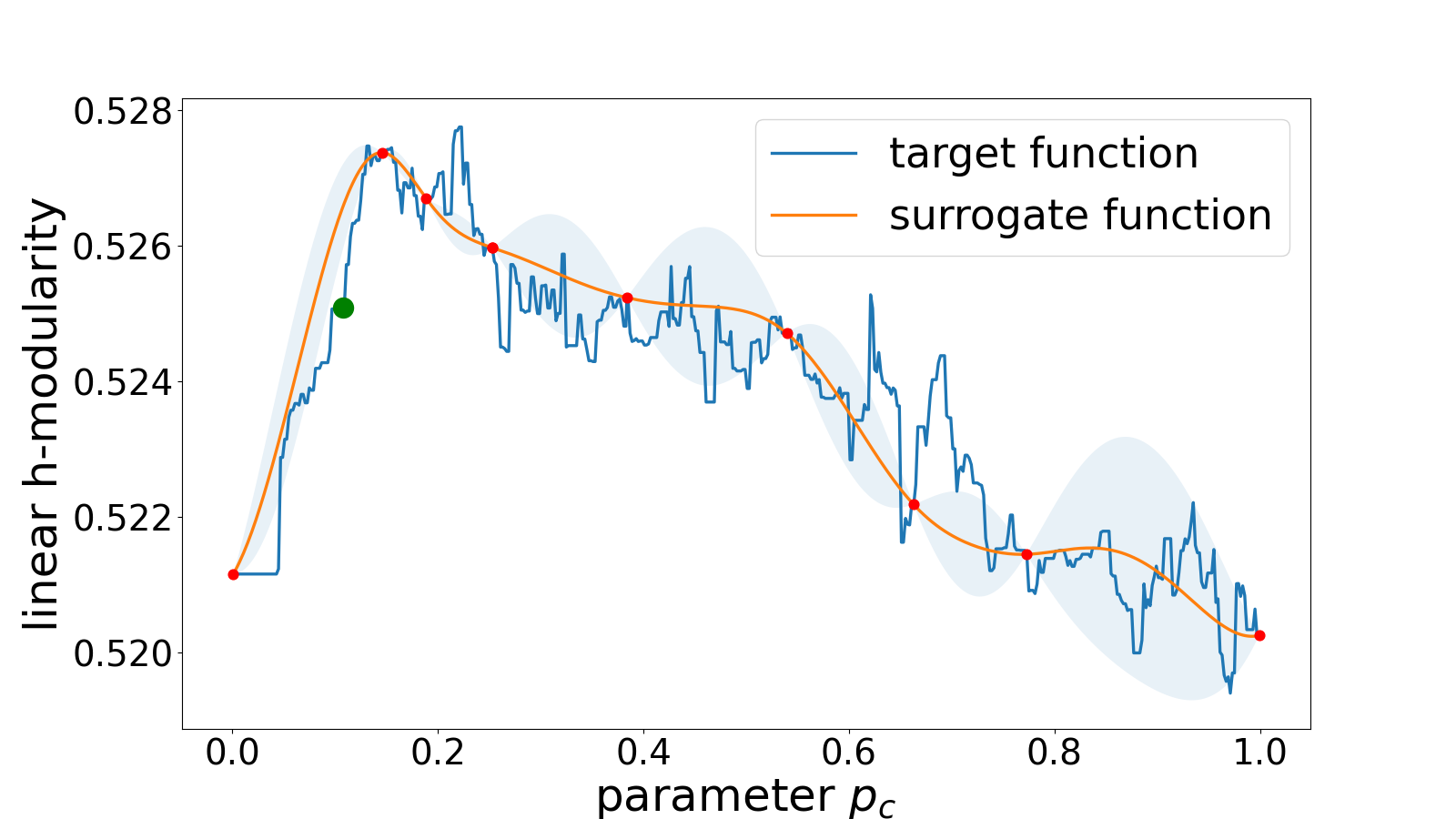}%
            \label{subfig:c}%
        }\hfill
        \subfloat[Step 11]{%
            \includegraphics[width=.53\linewidth]{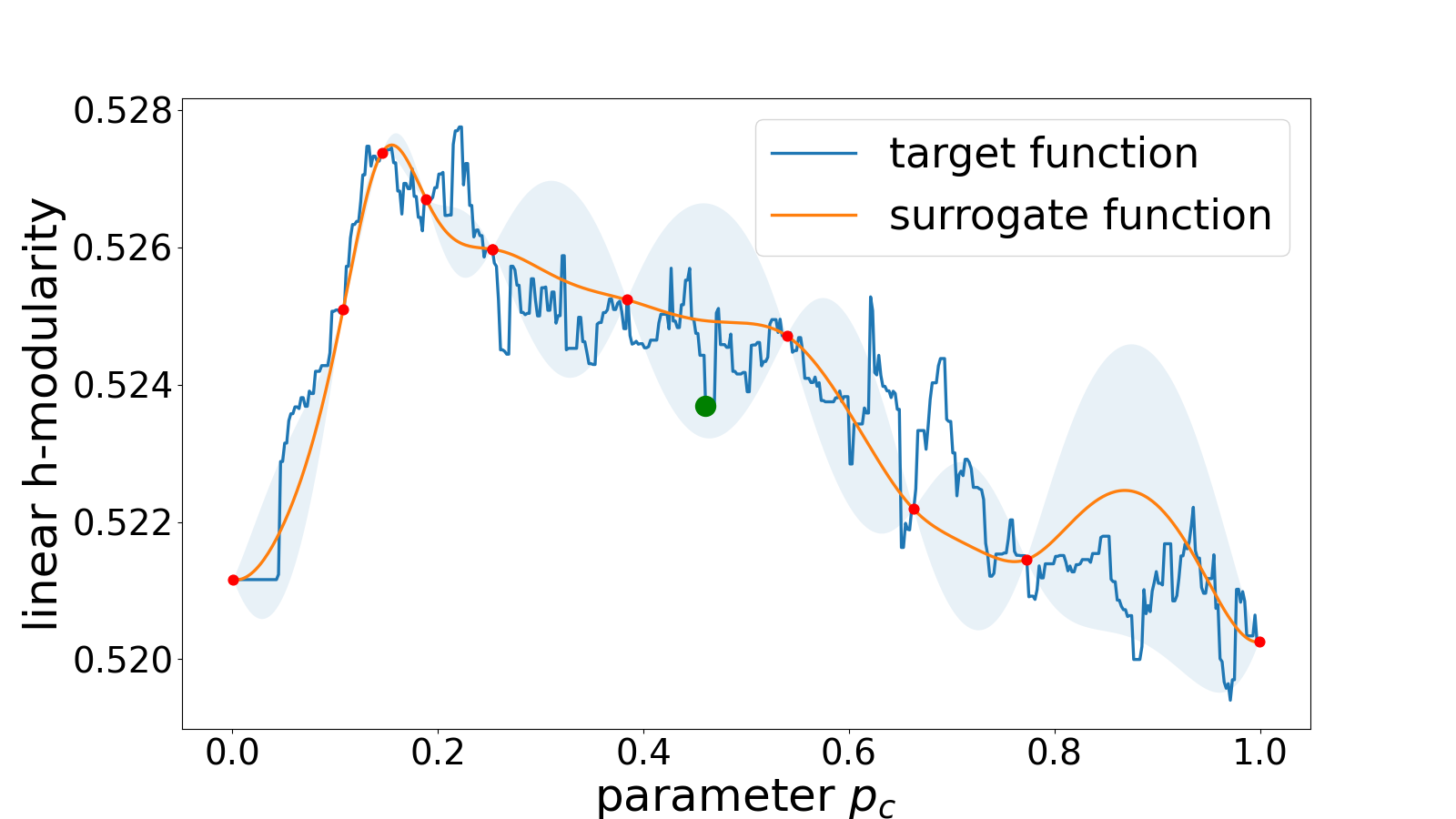}%
            \label{subfig:d}%
        }
        \caption{Visualization of the Bayesian optimization approach optimizing the modularity function returned by the \textbf{h--Louvain} algorithm.}
        \label{fig:BO}
    \end{figure*}

In order to visualize the Bayesian optimization procedure, we performed the following experiment. We selected one of the nine \textbf{h--ABCD} hypergraphs we experimented with (namely, the linear hypergraph with small level of noise, but this time with only $n=300$ nodes) 
and one of the three modularity functions (namely, the linear one) to be our target function. For cleaner visualization, we fixed $p_b = 0.9$ and used the procedure to find the optimum value of $p_c$ that maximizes the selected modularity function. Figure~\ref{fig:BO} presents situation at step $k$ of the algorithm for $k \in \{8, 9, 10, 11\}$. The blue curve is the target function that we independently computed but it is not available to the Bayesian optimization. The orange curve is a surrogate for the target function based on $k-1$ sampled points that are marked on this curve. The level of uncertainty is represented by the shaded area around this curve. Based on this information, the Bayesian optimization selects the next point to sample at this step which is depicted as a green point that lies outside of the orange curve. Note that the blue curve is deterministic (as we use fixed seeding of random number generator), as discussed above. It still has visible local variability, although it is possible to identify the region of good values of the parameter $p_c$ (in this case around $0.2$). This variability is expected and is the reason why when computing it we take the average of 10 independent evaluations of the algorithm (this approach significantly reduces the level of noise).

%%%%%%%%%%%%%%%%%%%%%%%%%%%%%%%%%%%%%%%%%%%%%%%%%%%%%%%%%%%
\section{Experiments}\label{sec:experiments}
%%%%%%%%%%%%%%%%%%%%%%%%%%%%%%%%%%%%%%%%%%%%%%%%%%%%%%%%%%%

In this section, we present several experiments aimed at testing our \textbf{h--Louvain} clustering algorithm as well as comparing outcomes of selecting various hypergraph modularity functions. One general observation is that the choice of the objective function to optimize, here the hypergraph $\tau$-modularity, typically has an enormous impact on the quality of the results (see Subsection~\ref{sec:modularity_selection}). Fortunately, one should be able to make a reasonable selection of a good value of $\tau$ in an unsupervised way. 

In general, in most of our experiments, we compare results obtained with our \textbf{h--Louvain} algorithm with results obtained by classical \textbf{Louvain} algorithm on the corresponding (weighted) 2-section graphs, as well as results using \textbf{Kumar}'s algorithm; see~\cite{Kumar1}. \textbf{Kumar}'s algorithm is a modification of the \textbf{Louvain} algorithm on 2-section graphs in which edges are re-weighted by taking into account the underlying hypergraph structure, and the hyperedge composition (with respect to the communities).

We consider synthetic hypergraphs with community structures, obtained via the \textbf{h--ABCD} benchmark, as well as four real-life hypergraphs, all of them described earlier in Section~\ref{sec:hypergraphs}. Synthetic hypergraphs allow us to investigate the performance of algorithms in various scenarios (see Subsection~\ref{sec:experiments_h-ABCD}), from hypergraphs with low level of noise ($\xi$ close to 0) that are easy to deal with to noisy hypergraphs ($\xi$ close to 1) that are challenging to find communities in. We also investigate a challenging case in which there are many hyperedges between two small communities (see Subsection~\ref{sec:more_challenging_ABCD}). It is known that many networks exhibit self-similar, ``fractal-type'' structure (see~\cite{barrett2023abcd,barrett2024self} and references therein) so such example aims to reproduce typical scenarios. This example highlights the power of our \textbf{h--Louvain} algorithm that in this particular setting clearly outperforms its competitors.

\medskip

For the real hypergraphs, we additionally consider the ``all of nothing'' (\textbf{AON}) variant of the \textbf{Louvain} algorithm. Specifically, we consider the version aiming to optimize the strict modularity, referred to as \textbf{AON}, as described in the associated GitHub repository\footnote{\url{https://github.com/nveldt/HyperModularity.jl}}. Note that, unless we start from some non-trivial partition such as the one obtained from the 2-section graph with \textbf{Louvain}, this algorithm requires 2-edges to be present. This is the case for the real hypergraphs we considered (but not so for several \textbf{h--ABCD} benchmark hypergraphs).

In general, the experiments on real-world hypergraphs (see Subsection~\ref{sec:exp-real_hypergraphs}) show that for appropriate value of $\tau$ affecting the choice of the objective function to optimize, one can improve the quality of the clusters measured by the \textbf{AMI} score with respect to the ground-truth. 
%\ft{I commented out the reference to $\gamma$ as I dropped the walmart example.}
%We also show that, similarly to what is known for graphs, one may tune the resolution parameter $\gamma$ in the modularity function, to recover the entire mesoscale of the hypergraph, from the trivial partition in which all nodes belong to the same community to another trivial partition in which every node forms its own community, hence the name multiresolution methods. 

%%%%%%%%%%%%%%%%%%%%%%%%%%%%%%%%%%%%%%%%%%%%%%%%%%%%%%%%%%%
\subsection{Selecting the Modularity Function to Optimize}\label{sec:modularity_selection}
%%%%%%%%%%%%%%%%%%%%%%%%%%%%%%%%%%%%%%%%%%%%%%%%%%%%%%%%%%%

Selecting an appropriate hypergraph modularity function to optimize is an important part of the process. The choice depends on how strongly one believes that a hyperedge is an indicator that at least some fraction of its nodes fall into one of the communities. In some situations, a reasonable assumption could be that not necessarily all members of that hyperedge must be in a single community but majority should (in such situations, quadratic modularity function might work well). On the other hand, some situations might have some underlying physical constraints that make one believe that all members should belong to one community unless such hyperedge is simply a noise (this time, strict modularity might be the one to optimize). If an analyst has some reasonable assumptions about the underlying process that created a hypergraph, then the decision which modularity function to use should be made based on this expert knowledge. In this section, we provide a general strategy for selecting a modularity function if such expert knowledge is not available, based on the structure of the hypergraph that can be detected in an unsupervised way.

Let us first start with highlighting important implications of the choice of the modularity function one decides to optimize. Recall that a community hyperedge (hyperedge with more than 50\% of members from one of the communities) of size $d$ that have exactly $c$ members from one of the communities is said to be of type $(c,d)$. In the absence of having the ground-truth available, one way to compare partitions returned by the algorithm aiming to optimize different modularity functions is to look at the distribution of edges of certain types. In our first experiment, we consider a synthetic \textbf{h--ABCD} graphs with only edges of size 5 and generated with the strict or the majority model, {\tt strict\_5} and {\tt majority\_5} with noise parameters $\xi = 0.3$ and $\xi = 0.2$. We run this experiment to show how the $(c,d)$ hyperedge composition changes under various modularity functions optimized. We expect that there should be visible difference in this distribution between {\tt strict\_5} and {\tt majority\_5} graphs independent of the choice of the objective function for community detection.

In Table~\ref{tab:edges}, we compare the distribution of edge types for 5 different partitions, namely, (i) the ground-truth partition, and partitioned returned by (ii) 2-section \textbf{Louvain}, (iii) \textbf{h--Louvain} with $\tau=2$ (quadratic modularity), (iv) \textbf{h--Louvain} with $\tau = 3$ (cubic modularity), and (v) \textbf{h--Louvain} with $\tau \to \infty$ (strict modularity). We count the number of edges with 5, 4 or, respectively, 3 nodes in the most frequent community; the remaining edges are considered to be noise.

\begin{table}[ht]
\centering
\begin{tabular}{c|c|c|c|c|c}
\hline \hline
Majority & Ground-truth & \textbf{Louvain}  & \multicolumn{3}{c}{ \textbf{h--Louvain} }  \\
class size & communities & 2-section & $\tau = 2$ & $\tau = 3$ & $\tau \to \infty$ (strict) \\
\hline \hline
\multicolumn{6}{c}{Strict with noise parameter $\xi = 0.2$} \\
\hline
5 & 352 & 352 & 349 & 352 & 352 \\
4 & 36 & 36 & 41 & 36 & 36 \\
3 & 92 & 92 & 91 & 92 & 92 \\
$\le 2$ & 42 & 42 & 41 & 42 & 42 \\
\hline \hline
\multicolumn{6}{c}{Strict with noise parameter $\xi = 0.3$} \\
\hline
5 & 314 & 314 & 311 & 314 & 314 \\
4 & 30 & 30 & 35 & 30 & 30 \\
3 & 123 & 123 & 122 & 123 & 123 \\
$\le 2$ & 55 & 55 & 54 & 55 & 55 \\
\hline \hline
\multicolumn{6}{c}{Majority with noise parameter $\xi = 0.2$} \\
\hline
5 & 169 & 170 & 137 & 169 & 264 \\
4 & 175 & 165 & 171 & 174 & 154 \\
3 & 137 & 138 & 161 & 137 & 104 \\
$\le 2$ & 41 & 49 & 53 & 42 & 0 \\
\hline \hline
\multicolumn{6}{c}{Majority with noise parameter $\xi = 0.3$} \\
\hline
5 & 158 & 129 & 88 & 158 & 206 \\
4 & 145 & 140 & 148 & 144 & 147 \\
3 & 158 & 151 & 196 & 161 & 169 \\
$\le 2$ & 61 & 102 & 90 & 59 & 0 \\
\hline \hline
\end{tabular}
\vspace{.25cm}
\caption{The number of hyperedges of each type for \textbf{h--ABCD} hypergraphs with 5-edges generated with strict ({\tt strict\_5}) or majority ({\tt majority\_5}) assignment rule.}
\label{tab:edges}
\end{table}

In the second column of in Table~\ref{tab:edges} we show hyperedge composition of the ground truth. As expected, for hypergraph with the strict model used the $(5,5)$ hyperedges are most common, and for hypergraph with the majority model used $(5,5)$, $(4,5)$, and $(3,5)$ hyperedges have similar frequencies (this holds both for $\xi=0.2$ and $\xi=0.3$). The crucial observation is that regardless of which of the modularity function is optimized (\textbf{Louvain} 2-section or \textbf{h--Louvain} with varying $\tau$), the recovered hyperedge composition is similar to the ground truth. This observation suggests using the following approach in cases where the user does not have a prior preference for the $\tau$ parameter in the \textbf{h--Louvain} algorithm.

As a rule of thumb, running a quick clustering (for example with 2-section \textbf{Louvain}) as a part of Exploratory Data Analysis (EDA), and looking at the composition of edge types is a recommended first step that can be used to decide on the value(s) of $\tau$ one wants to use as the objective $\tau$-modularity function for \textbf{h--Louvain}. In general, there are two major possible scenarios that the user could consider. Seeing mostly ``'pure'' edges suggests using large value of $\tau$ (or strict modularity), while the opposite suggests using a smaller values of $\tau$ such as $\tau = 2$ or $\tau=3$.

%%%%%%%%%%%%%%%%%%%%%%%%%%%%%%%%%%%%%%%%%%%%%%%%%%%%%%%%%%%
\subsection{Synthetic \textbf{h--ABCD} Hypergraphs}\label{sec:experiments_h-ABCD}
%%%%%%%%%%%%%%%%%%%%%%%%%%%%%%%%%%%%%%%%%%%%%%%%%%%%%%%%%%%

We ran a series of experiments using the synthetic \textbf{h--ABCD} benchmark hypergraphs.
For each family of hypergraphs, we considered a wide range of values for the noise parameter $\xi$, and for each $\xi$, we generated 30 independent copies of \textbf{h--ABCD} hypergraphs. For each hypergraph, we obtained clusterings in various ways:
\begin{itemize}
    \item taking the 2-section (weighted) graph and applying the \textbf{Louvain} algorithm several times, keeping the partition with the largest (graph) modularity;
    \item running \textbf{Kumar}'s algorithm;
    \item running our \textbf{h--Louvain} algorithm with Bayesian optimization for $\tau=2$ and $\tau=3$, and
    \item running our \textbf{h--Louvain} algorithm with Bayesian optimization using the strict modularity ($\tau \to \infty$) as the objective function.
\end{itemize}

\begin{figure*}[ht]
\centering
\includegraphics[width=.45\linewidth]{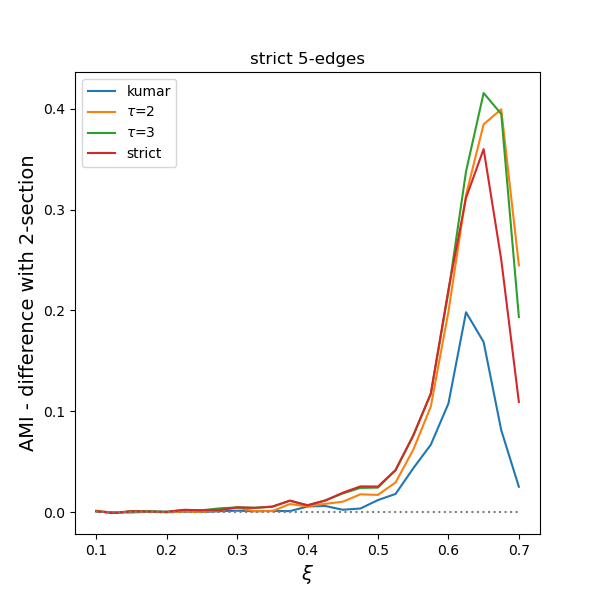}
\hspace{.1cm}
\includegraphics[width=.45\linewidth]{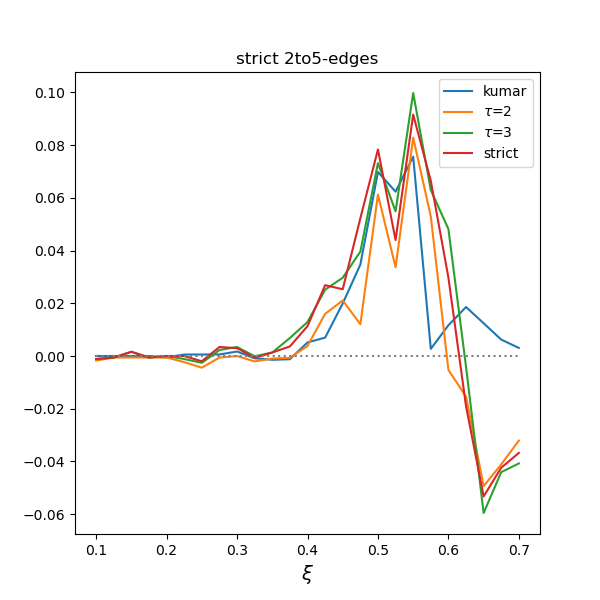}
\caption{Results with \textbf{h--ABCD} hypergraphs with strict model for the community edge composition, ({\tt strict\_5} and {\tt strict\_2to5}) showing \textbf{AMI} difference between 2-section communities and the considered algorithms. Positive values indicate increase of \textbf{AMI} for a given algorithm.}
\label{fig:large_strict}
\end{figure*}

\begin{figure*}[ht]
\centering
\includegraphics[width=.45\linewidth]{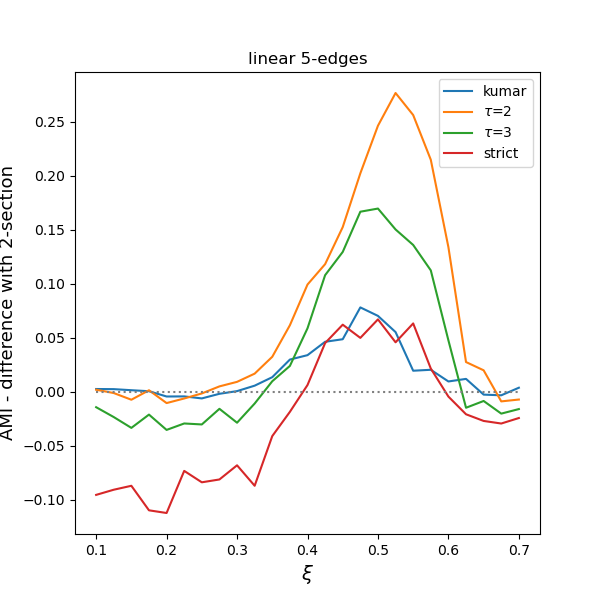}
\includegraphics[width=.45\linewidth]{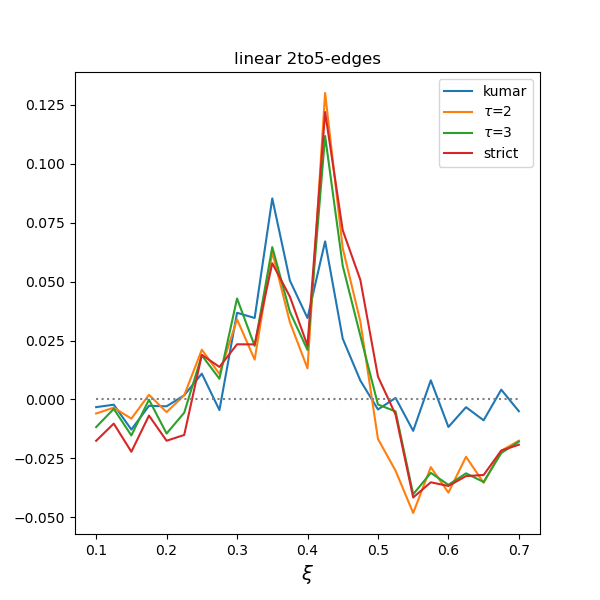}
\caption{Results with \textbf{h--ABCD} hypergraphs with linear model for the community edge composition, ({\tt linear\_5} and {\tt linear\_2to5}) showing \textbf{AMI} difference between 2-section communities and the considered algorithms. Positive values indicate increase of \textbf{AMI} for a given algorithm.}
\label{fig:large_linear}
\end{figure*}

\begin{figure*}[ht]
\centering
\includegraphics[width=.45\linewidth]{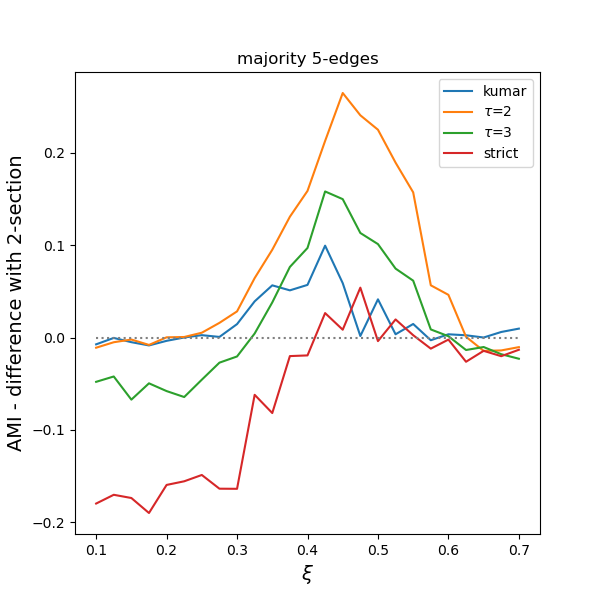}
\includegraphics[width=.45\linewidth]{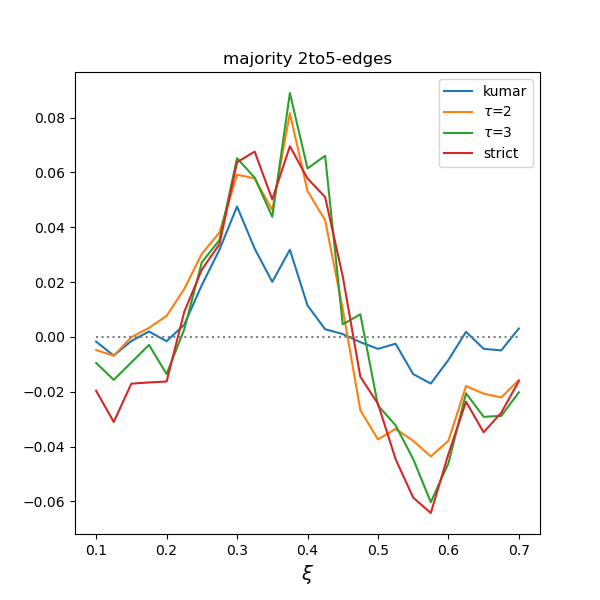}
\caption{Results with \textbf{h--ABCD} hypergraphs with majority model for the community edge composition, ({\tt majority\_5} and {\tt majority\_2to5}) showing \textbf{AMI} difference between 2-section communities and the considered algorithms. Positive values indicate increase of \textbf{AMI} for a given algorithm.}
\label{fig:large_majority}
\end{figure*}

In the analysis of the results, we computed the \textbf{AMI} of each partition with respect to the ground truth communities. The plots in Figures~\ref{fig:large_strict}--\ref{fig:large_majority} show the difference of the \textbf{AMI} of a given algorithm and the \textbf{AMI} of 2-section result. In other words, we measure how much gain/loss is obtained by switching from a standard 2-section approach to finding communities in a hypergraph to our algorithm designed specifically for hypergraphs.

\clearpage

Here are some general remarks from those experiments:
\begin{itemize}
    \item The hypergraph specialized (\textbf{h-Louvain} and \textbf{Kumar}) algorithms give the most substantial benefits for moderately noisy hypergraphs. The reason is that for hypergraphs with very low level of noise (values of $\xi$ close to zero) all algorithms produce similar results, as the community-finding problem is simple, and for very noisy graph (values of $\xi$ close to one) the noise itself creates spurious communities that the algorithm start to recover (this effect has been previously studied and analytically analyzed for the \textbf{ABCD} graphs in~\cite{kaminski2022modularity}).
    \item Our \textbf{h--Louvain} algorithm with $\tau=2$ and $\tau=3$ outperforms \textbf{Kumar}'s algorithm most of the time. The exceptions are a few cases with large amount of noise in the hypergraph (values of $\xi$ close to one).
    \item Strict \textbf{h--Louvain} modularity function ($\tau\to\infty$) may work poorly for hypergraphs that have many non-pure community hyperedges. For this reason, in the case of absence of a prior preference, users should follow the initial verification procedure described in Section~\ref{sec:modularity_selection} before using this parameterization. The reason is that for $\tau\to\infty$, all hyperedges of type $(c,d)$ with $c<d$ are not counted as community hyperedges, which would loose potentially valuable information in cases when they are indeed informative.
\end{itemize}

%%%%%%%%%%%%%%%%%%%%%%%%%%%%%%%%%%%%%%%%%%%%%%%%%%%%%%%%%%%
\subsection{More Challenging Case---Synthetic Hypergraphs with Localized Noise}\label{sec:more_challenging_ABCD}

In this experiment, we simulate an example in which the difficulty in recovering communities is due to the fact that several ``noise'' edges touch a small number of communities instead of being sprinkled over several communities. To that end, we generated the \textbf{h--ABCD} hypergraph with $n = 300$ nodes, degree exponent $\alpha = 2.5$ in the range $[5,30]$, community size exponent $\beta = 1.5$ in the range $[40,60]$, edges of size 5 with purity distribution for community hyperedges set to (0.7, 0.2, 0.1), i.e.\ 70\% of them have 3 community nodes (type $(3,5)$), 20\% have 4 (type $(4,5)$), and 10\% have 5 (pure hyperedges, type $(5,5)$). Overall noise is set to $\xi = 0.2$, but we also add 35 additional 5-edges where nodes are randomly sampled from the two smallest communities. This simulates ``localized'' noise which should make the community detection more challenging.

First, simulating a real-life application of the procedure we proposed in Subsection~\ref{sec:modularity_selection}, we look at the edge composition when running a 2-section clustering, which is reported in Table~\ref{tab:habcd_compo}. Running this quick experiment is indicative that smaller values for $\tau$ are likely to be a better choice than using the strict modularity version, since there are not that many ``pure'' edges.

\begin{table}[ht]
\centering
\begin{tabular}{rrr}
\toprule
$d$ & $c$ & frequency \\
\midrule
5 & 5 & 58 \\
5 & 4 & 158 \\
5 & 3 & 247 \\
5 & 2 & 120 \\
5 & 1 & 7 \\
\bottomrule
\end{tabular}
\vspace{.25cm}
\caption{Number of edges of each type for \textbf{h--ABCD} hypergraphs with 5-edges and localized noise added. The partition was obtained by running \textbf{Louvain} on the weighted 2-section graph.}
\label{tab:habcd_compo}
\end{table}

\begin{figure*}[ht]
\centering
\includegraphics[width=.75\linewidth]{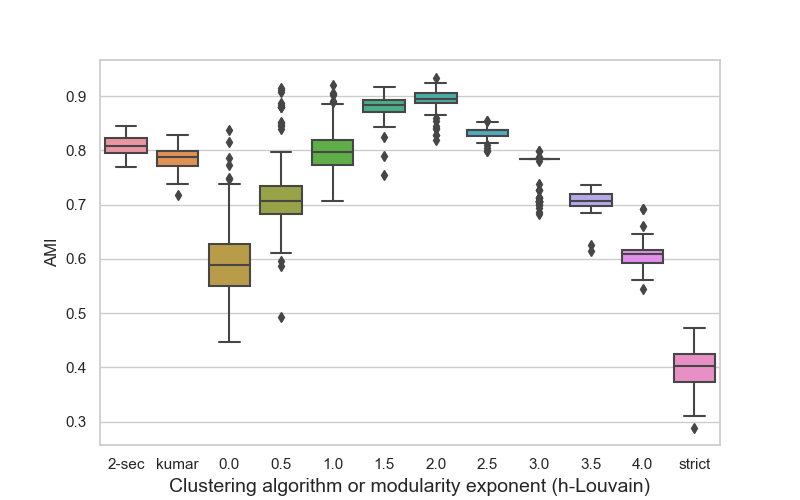}%
\caption{Results of 100 runs for several choices of hypergraph $\tau$-modularity for \textbf{h--Louvain} (strict, and with $0 \le \tau \le 4$) as well as using \textbf{Louvain} on 2-section modularity and \textbf{Kumar}.}
\label{fig:habcd_local}
\end{figure*}

We did 100 runs for each choice of the $\tau$-modularity for our \textbf{h--Louvain}: strict ($\tau \to \infty$) and $\tau \in \{0, 0.5, 1, 1.5, 2, 2.5, 3, 3.5, 4\}$. We also did 100 runs using the 2-section modularity with \textbf{Louvain}, and \textbf{Kumar}'s algorithm. The results are presented in Figure~\ref{fig:habcd_local}.
From those results, we see that clustering the 2-section graph or using \textbf{Kumar}'s algorithm yield good results, but we can improve those results when optimizing the hypergraph $\tau$-modularity when choosing $\tau \approx 2$. As expected from the preliminary EDA analysis we performed earlier, using small values for $\tau$ (close to zero) or large values for $\tau$ (including strict modularity, $\tau \to \infty$) are bad choices in this case. 

%%%%%%%%%%%%%%%%%%%%%%%%%%%%%%%%%%%%%%%%%%%%%%%%%%%%%%%%%%%
\subsection{Real-world Hypergraphs}
\label{sec:exp-real_hypergraphs}

We consider two real-world hypergraphs the {\tt primary-school} contact and the {\tt cora}. We first run the EDA procedure, suggested in Subsection~\ref{sec:modularity_selection}, on both graphs by looking at the hyperedge composition associated with the corresponding 2-section clusterings. The results are presented in Table~\ref{tab:emp_2sec}. We can see that the {\tt primary-school} hypergraph (left) has relatively more non-pure hyperedges than the {\tt cora} hypergraph. This indicates that one should expect that for the {\tt primary-school} case the optimal $\tau$ is smaller than the one for the {\tt cora} hypergraph.

\begin{table}[ht]
\centering
\begin{tabular}{ccrc}
\multicolumn{4}{c}{\tt primary-school} \\

\toprule
d & c & purity & frequency \\
\midrule
2 & 1 & 50\% & 4051 \\
\rowcolor{lightgray}
2 & 2 & 100\% & 3697  \\
\rowcolor{lightgray}
3 & 3 & 100\% & 3385 \\
\rowcolor{lightgray}
3 & 2 & 67\% & 1054 \\
3 & 1 & 33\% & 161 \\
\rowcolor{lightgray}
4 & 4 & 100\% & 240  \\
\rowcolor{lightgray}
4 & 3 & 75\% & 58 \\
4 & 2 & 50\% & 47 \\
\rowcolor{lightgray}
5 & 4 & 80\% & 3 \\
5 & 3 & 60\% & 3 \\
\bottomrule
\end{tabular}
\hspace{1.0cm}
\begin{tabular}{ccrc}
\multicolumn{4}{c}{\tt cora} \\
\toprule
d & c & purity & frequency \\
\midrule
\rowcolor{lightgray}
2 & 2 & 100\% & 512 \\
\rowcolor{lightgray}
3 & 3 & 100\% & 354 \\
\rowcolor{lightgray}
4 & 4 & 100\% & 212 \\
\rowcolor{lightgray}
5 & 5 & 100\% & 108 \\
\rowcolor{lightgray}
3 & 2 & 67\% & 85 \\
\rowcolor{lightgray}
4 & 3 & 75\% & 62 \\
2 & 1 & 50\% & 60 \\
\rowcolor{lightgray}
5 & 4 & 80\% & 41 \\
4 & 2 & 50\% & 32 \\
\rowcolor{lightgray}
5 & 3 & 60\% & 21 \\
\bottomrule
\end{tabular}
\vspace{.25cm}
\caption{The number of edges of each type (top-10 most frequent) for the {\tt primary-school} contact (left) and {\tt cora} (right) hypergraphs. The corresponding partitions were obtained by running \textbf{Louvain} on the weighted 2-section graph.}
\label{tab:emp_2sec}

\end{table}

%%%%%%%%%%%%%%%%%%%%%%%%%%%%%%%%%%%%%%%%%%%%%%%%%%%%%%%%%%%
\subsubsection{Contact Hypergraph}

\begin{figure*}[ht]
\centering
\includegraphics[width=.75\linewidth]{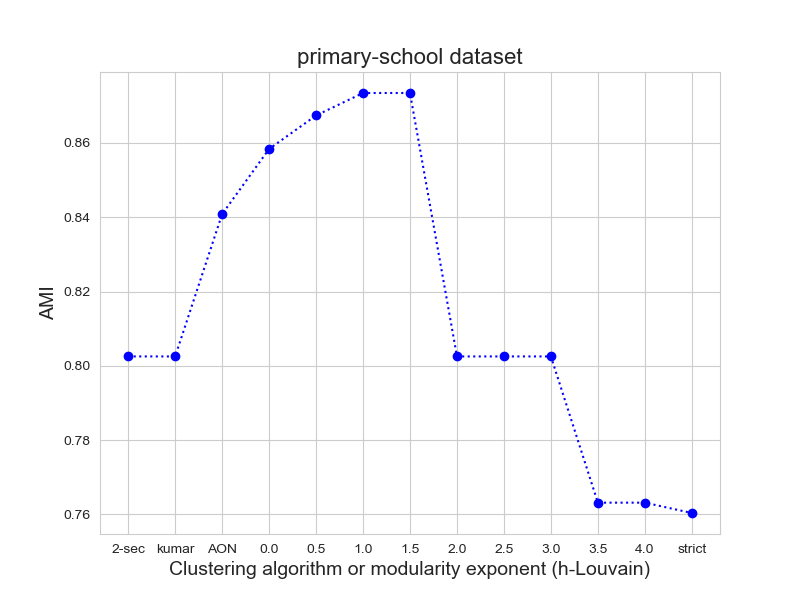}%
\caption{Results for several choices of hypergraph $\tau$-modularity for \textbf{h--Louvain} (strict and with $0 \le \tau \le 4$) as well as using 2-section modularity and \textbf{Louvain}, \textbf{Kumar}'s, and \textbf{AON} algorithms.}
\label{fig:school}
\end{figure*}

Let us first consider the {\tt primary-school} contact hypergraph described in Section~\ref{sec:real_hypergraphs}.
The results are shown in Figure~\ref{fig:school}, where we compare 2-section (graph) clustering with \textbf{Louvain}, \textbf{Kumar}, and \textbf{AON} clustering as well as our \textbf{h--Louvain} using different values of $\tau$ for the modularity function. The \textbf{AMI} scores are averaged over 30 runs. The variance is negligible and is not shown. 
From this experiment, we see that one can get some improvement over 2-section and \textbf{Louvain} or \textbf{Kumar}'s algorithm when using small values for $\tau$ in our \textbf{h--Louvain}.

\subsubsection{Co-citation Hypergraphs}

Next, we consider the {\tt cora} co-citation hypergraph described in Subsection~\ref{sec:real_hypergraphs} in which nodes are publications which belong to 7 categories, and hyperedges represent co-citations. Since the graph has several small disconnected components, we restrict ourselves to the giant component which has 1,330 nodes and 1,503 hyperedges. 

\begin{figure*}[ht]
\centering
\includegraphics[width=13cm]{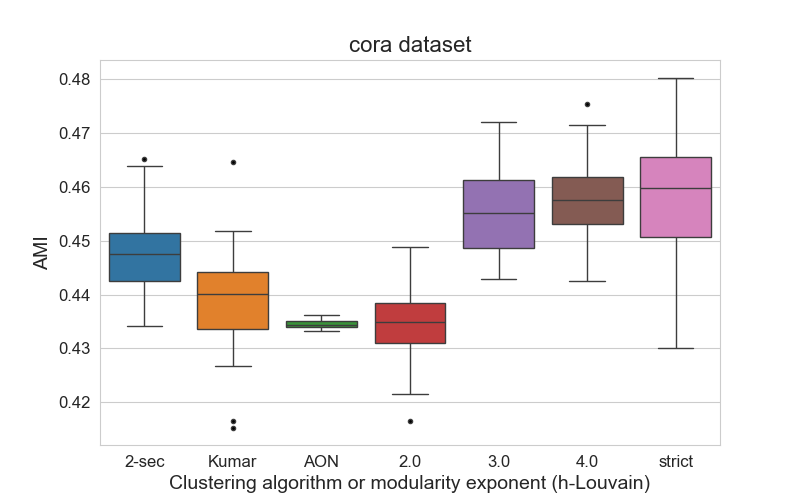}
\caption{Clustering the {\tt cora} co-citation hypergraph.}
\label{fig:cora}
\end{figure*}

We ran each clustering algorithm 50 times, with the results reported in Figure~\ref{fig:cora}. 
The results with \textbf{AON} were worse in this case (with \textbf{AMI} around 0.21, not reported in Figure~\ref{fig:cora}). Instead, we report the results when starting from a partition returned by the 2-section graph clustering before running \textbf{AON}, which give better results.
As expected from the EDA analysis comparing {\tt primary-school} and {\tt cora} hypergraphs, for the {\tt cora} hypergraph, we get good results running \textbf{h--Louvain} with values of $\tau$ larger than for the {\tt primary-school} hypergraph, slightly improving on the results with 2-section graph clustering with \textbf{Louvain}, \textbf{Kumar}'s algorithm, or \textbf{AON}.

%%%%%%%%%%%%%%%%%%%%%%%%%%%%%%%%%%%%%%%%%%%%%%%%%%%%%%%%%%%
\section{Conclusions}\label{sec:conclusions}
%%%%%%%%%%%%%%%%%%%%%%%%%%%%%%%%%%%%%%%%%%%%%%%%%%%%%%%%%%%

In this paper, we proposed a modification of the classical \textbf{Louvain} algorithm that allows us to optimize the hypergraph modularity, \textbf{h-Louvain}. Our approach is to optimize a weighted average of the 2-section graph modularity and the hypergraph modularity, with an increasing weight of hypergraph modularity component as the optimization process progresses.
We presented both theoretical arguments as well as empirical evidence that the approach of increasing the weight of the hypergraph modularity component is efficient. 
Since there are several ways to update this weight,
we developed a method allowing for automatic selection of hyperparameters of this process using Bayesian optimization. 
We have shown that the {\bf h--Louvain} algorithm is competitive and, in particular, that it can outperform both {\bf Louvain} on 2-section graph and {\bf Kumar}'s algorithms in terms of recovering ground truth communities both for synthetic and real networks.

Additionally, let us mention about another important and interesting aspect. Since in \textbf{h--Louvain} the optimization process is stochastic by nature, the results of a single optimization pass can be easily improved by running many such optimizations in parallel. Therefore, an important extension to the algorithm is for allowing it to learn how to dynamically set the tuneable parameters when multiple optimization processes are executed.

\bibliography{ref}

\clearpage 

\appendix

\section{Appendix---Pseudo-code of the \textbf{h-Louvain} Algorithm}\label{apendix_code}

%% Change to Input/Output 
\algrenewcommand\algorithmicrequire{\textbf{Input:}}
\algrenewcommand\algorithmicensure{\textbf{Output:}}

\begin{algorithm}[ht]
\small
\label{alg:hlouvain}
	\caption{h-Louvain({\it{H}, \it{$\Gamma$}})} 
	\begin{algorithmic}[1]
	\Require $H = (V,E)$ -- input hypergraph; $\Gamma$ -- policy to control $\alpha \in [0,1]$ defined using parameters $p_b$ and $p_c$ 
    \Ensure ${\bf{A}}$ -- partition of $V$; $q_H(\bf{A})$ - hypergraph modularity 
    \State Initialize: Build $ G = (V, E_G)$ (2-section), and set partition $\bf{A}$ with all vertices $v\in V$ in their own cluster 
    %compute $q(A)$
    \State {\it modified} $\leftarrow$ True
    \State $\alpha \leftarrow 0$ % UpdateAlpha($\bf{A}$,$\Gamma$) \pmc{now it can be just $\alpha \leftarrow 0$}
    \State $n \leftarrow |V|$
    \While  {{\it modified}}
    \State {\it modified} $\leftarrow$ False
        \State {\it improved} $\leftarrow$ True
		\While  {{\it improved}}
		    \State {\it improved} $\leftarrow$ False
			\State randomize the order of vertices in $V$
			\For {$v \in V$} 
			    \State {\it bestCommunity} $\leftarrow A_i$ (current community of $v$)
			    \State $bestDelta \leftarrow 0$
			    \For {all neighbouring communities $A_j$ of $v$} 
			        \State $deltaModularity  \leftarrow (1-\alpha) \Delta q_G(A') + \alpha \Delta q_H(A')$ ($A'$: $v$ moved from $A_i$ to $A_j$)
			        \If{$deltaModularity > bestDelta$}
			            \State $bestDelta \leftarrow deltaModularity$
			            \State $bestCommunity \leftarrow$ $A_j$ 
			            \State $improved \leftarrow$ True
			        \EndIf
				\EndFor
				\If {$improved$}
				    \State change $\bf{A}$ by moving $v$ to $bestCommunity$
				    \State $modified \leftarrow$ True
                    \State $\alpha \leftarrow$ UpdateAlpha($\bf{A}$,$n$,$\alpha$,$\Gamma$)
				\EndIf
			\EndFor
		\EndWhile
		\If {$modified$}
		    \State update $H = (V,E)$ (merge current communities into supernodes and update edges)
        \ElsIf {$\alpha <1$}
            \State $\alpha \leftarrow 1$
            \State revert the last merging communities step        
            \State $modified \leftarrow$ True
		\EndIf
	\EndWhile
    \State return $\bf{A}$, $q_H(\bf{A})$
	\end{algorithmic} 
\end{algorithm}

\begin{algorithm}
\small
\label{alg:example_policy}
	\caption{UpdateAlpha({\bf{A}, $n$, $\alpha$, \it{$\Gamma$}})}  %\pmc{should we replace $\Gamma$ with $p_b$ and $p_c$?}
	\begin{algorithmic}[1]
	\Require current partition $\bf{A}$, total number of nodes $n$, previous value of $\alpha$, policy $\Gamma$ to control $\alpha \in [0,1]$ defined using parameters $p_b$ and $p_c$ 
    \Ensure new value of $\alpha$ 
    \If {$\alpha < 1$} %\pmc{Line 1: $\alpha$ is not procedure input, should it be $p_b$?}
        \State $|\bf{A}| \leftarrow$ number of communities in current partition $\bf{A}$
        \State $j \leftarrow \arg \max_{k \in \mathbb{N}}(k: \frac{|{\bf{A}}|}{n} \le p_c^{k})$ 
        \State $\alpha \leftarrow 1 - (1 - p_b)^{j}$
        \State return $\alpha$
    \Else 
        \State return 1
    \EndIf
    	\end{algorithmic} 
\end{algorithm}

\end{document}